\documentclass[12pt]{iopart}
\usepackage{iopams}  
\usepackage{revsymb}
\usepackage{graphicx}
\usepackage{lscape}
\begin{document}

\newcommand{\apj}{{Astrophys.\ J. }}
\newcommand{\apjs}{{Astrophys.\ J.\ Supp. }}
\newcommand{\apjl}{{Astrophys.\ J.\ Lett. }}
\newcommand{\aj}{{Astron.\ J. }}
\newcommand{\prl}{{Phys.\ Rev.\ Lett. }}
\newcommand{\prd}{{Phys.\ Rev.\ D }}
\newcommand{\mnras}{{Mon.\ Not.\ R.\ Astron.\ Soc. }}
\newcommand{\araa}{{ARA\&A }}
\newcommand{\aap}{{Astron.\ \& Astrophy. }}
\newcommand{\nat}{{Nature }}
\newcommand{\cqg}{{Class.\ Quantum Grav.\ }}

\def\HS{{\mathfrak H}_3}
\def\bfis{\hbox{\scriptsize\rm i}}
\def\bfi{\hbox{\rm i}}
\def\bfj{\hbox{\rm j}}
\def\3{{\ss}}

\hfill astro-ph/0612308

\title[CMB Alignment in Multi-Connected Universes]
{CMB Alignment in Multi-Connected Universes}

\addtocounter{footnote}{-1}
\author{Ralf Aurich, Sven Lustig, Frank Steiner, and
Holger Then\footnotemark[1]$^1$}

\address{Institut f\"ur Theoretische Physik, Universit\"at Ulm,\\
Albert-Einstein-Allee 11, D-89069 Ulm, Germany}

\footnotetext[1]{$^1$ Present address:
Carl-von-Ossietzky Universit\"at Oldenburg,
Institut f\"ur Physik, D-26111 Oldenburg, Germany}

\begin{abstract}
The low multipoles of the cosmic microwave background (CMB) anisotropy
possess some strange properties like the alignment of the
quadrupole and the octopole, and the extreme planarity or the extreme
sphericity of some multipoles, respectively.
In this paper the CMB anisotropy of several multi-connected space forms
is investigated with respect to the maximal angular momentum dispersion
and the Maxwellian multipole vectors in order to settle the question
whether such spaces can explain the low multipole anomalies in the CMB.
\end{abstract}





\section{Introduction}

The study of the properties of the low multipoles of the
cosmic microwave background (CMB) anisotropy requires full-sky maps
as provided by the Wilkinson Microwave Anisotropy Probe (WMAP)
\cite{Bennett_et_al_2003,Hinshaw_et_al_2006}.
These large-scale studies allow to scrutinise the assumptions
on statistical isotropy and Gaussianity as predicted by
inflationary scenarios. 
Such an analysis requires a foreground-cleaned map
as the Internal Linear Combination (ILC) map \cite{Bennett_et_al_2003b}
and the ``TdOH'' maps derived by
\cite{Tegmark_deOliveira_Costa_Hamilton_2003}.
The best known anomalies at low values of $l$ are the quadrupole suppression
and the very small values of the two-point temperature correlation function
$C(\vartheta)$ at angular separations $\vartheta \gtrsim 60^\circ$.
This low power at angular scales above $60^\circ$ can be well explained
by assuming a multi-connected space for the spatial sections
of our Universe.
Besides this quadrupole suppression,
an anomalous alignment between the quadrupole and
the octopole was discovered in
\cite{Tegmark_deOliveira_Costa_Hamilton_2003,%
deOliveira-Costa_Tegmark_Zaldarriaga_Hamilton_2004},
the latter being unusually planar.
This alignment is described by the axis $\hat n_l$ which maximises
the {\it angular momentum dispersion}
\cite{Tegmark_deOliveira_Costa_Hamilton_2003}
\begin{equation}
\label{Eq:AMD}
(\Delta L(\hat n))^2_l \; := \; \sum_{m=-l}^l m^2 \, |a_{lm}(\hat n)|^2
\end{equation}
for a given multipole $l$.
Here $a_{lm}(\hat n)$ denotes the expansion coefficients of 
the CMB temperature fluctuation $\delta T$ with respect to
the spherical harmonics $Y_{lm}(\vartheta,\phi)$
with the $z$-axis pointing in the direction of $\hat n$.
This prescription gives for each multipole $l$ only one unit vector $\hat n_l$
thus leading to a loss of information.

An alternative statistics with the aim to analyse the multipole alignment
is given by \cite{Land_Magueijo_2005b}
\begin{equation}
\label{Eq:rl_statistic}
r_l \; := \; \max_{m,\hat n} \,
\frac{{\cal C}_{lm}(\hat n)}{(2l+1) C_l}
\end{equation}
with ${\cal C}_{l0}(\hat n) = |a_{l0}(\hat n)|^2$,
${\cal C}_{lm}(\hat n) = 2 |a_{lm}(\hat n)|^2$ for $m>0$ and
$(2l+1) C_l = \sum_{m=-l}^l |a_{lm}|^2$.
This statistics selects the direction $\hat n$
which concentrates the most power into a single $m$ mode.
In contrast to (\ref{Eq:AMD}) the statistics (\ref{Eq:rl_statistic})
does not prefer the $m$ mode with $m=l$.
We find in agreement with \cite{Copi_Huterer_Schwarz_Starkman_2005}
that this statistics is very sensitive to noise which changes the
selected value of $m$ and in turn leads to a completely
different direction $\hat n$.
Nevertheless the anomalous multipole alignment
is confirmed using (\ref{Eq:rl_statistic}) in \cite{Land_Magueijo_2005b}.

Instead of the axis $\hat n_l$ of maximal angular momentum dispersion
(\ref{Eq:AMD}), one can use multipole vectors \cite{Maxwell_1891}.
The expansion of a function $f(\vartheta,\phi)$ on the sphere ${\cal S}^2$
\begin{equation}
\label{Eq:multipole_expansion}
f(\vartheta,\phi) \; = \; \sum_{l=0}^\infty f_l(\vartheta,\phi)
\end{equation}
is provided by the Maxwellian multipole vectors
$\hat v^{(l,j)}$, $j=1,\dots,l$, and a scalar $A^{(l)}$ with
\begin{equation}
\label{Eq:Maxwell_multipole_expansion}
f_l(\vartheta,\phi) = \left[ A^{(l)} \; (\hat v^{(l,1)} \cdot \vec \nabla)
\, \cdots \, (\hat v^{(l,l)} \cdot \vec \nabla) \; \frac 1r \right]_{r=1}
\hspace{10pt}.
\end{equation}
The multipole vectors $\hat v^{(l,j)}$ together with $A^{(l)}$ contain
the complete information about $f(\vartheta,\phi)$ in the same way
as the $a_{lm}$'s of the usual spherical harmonics expansion.
For $l=2$ there is the relation
$\hat n_2 = \pm \, \hat v^{(2,1)} \times  \hat v^{(2,2)}$,
however for $l>2$ there does not exist such a simple relation
\cite{Copi_Huterer_Schwarz_Starkman_2005}.

The multipole vectors have been applied to the CMB anisotropy
in \cite{Copi_Huterer_Starkman_2004, Schwarz_Starkman_Huterer_Copi_2004}
in order to study the large-angle anomalies.
Area vectors
\begin{equation}
\label{Eq:Area_vectors}
\vec w^{(l,i,j)} := \pm \, \hat v^{(l,i)} \times  \hat v^{(l,j)}
\end{equation}
are introduced which give the normals to the planes defined by
each pair of multipole vectors.
The signs are arbitrary and chosen such that the area vectors
$\vec w^{(l,i,j)}$ point towards the northern galactic hemisphere.
The quadrupole plane and the three octopole planes are found to be
strongly aligned \cite{Schwarz_Starkman_Huterer_Copi_2004}.
Furthermore, three of these four planes are found to be orthogonal to
the ecliptic.
This result might point to a neglected solar foreground.
Correcting this supposed foreground might resolve the mysterious
low-$l$ anomalies.

Assuming that there are no unknown foregrounds and
no systematic errors in the measurements,
the low-$l$ anomalies have to be considered as of cosmological origin.
One such possibility is that the statistical isotropy is violated
by vorticity and shear contributions in the metric due to a
Bianchi VII$_{\hbox{\scriptsize h}}$ cosmology.
Such a model could explain the anomalies but for an unrealistically
low value of $\Omega_{\hbox{\scriptsize tot}} \simeq 0.5$
for a pure matter model
\cite{Jaffe_Banday_Eriksen_Gorski_Hansen_2005,Land_Magueijo_2005a}.
Even the inclusion of dark energy does not lead to more viable
cosmological parameters in order to explain the anomalies
\cite{Jaffe_Hervik_Banday_Gorski_2005}.
The anomalies could also be caused by matter inhomogeneities of the
local Universe ($z \lesssim 1$) as an asymmetric distribution of voids
\cite{Inoue_Silk_2006} using the usual cosmological parameters,
or the quadrupole could be modified by the Local Supercluster
via the Sunyaev-Zel'dovich effect \cite{Abramo_Sodre_Wuensche_2006_a}.
A further possible explanation could be
that the quadrupole-octopole alignment is rather an anti-alignment
of the quadrupole and the octopole with the dipole.
A slightly erroneous treatment of the large dipole could then lead
to a spurious quadrupole-octopole alignment
\cite{Helling_Schupp_Tesileanu_2006}.

An alternative explanation for the anomalies might come from cosmic topology,
where one assumes that the spatial section of the Universe is multi-connected.
In \cite{deOliveira-Costa_Tegmark_Zaldarriaga_Hamilton_2004}
toroidal space forms in a flat universe are considered
in which one side of the torus has a length below the Hubble radius
whereas the other two are above.
This so-called slab topology can explain the planarity of the octopole
but not the alignment between quadrupole and octopole
according to \cite{deOliveira-Costa_Tegmark_Zaldarriaga_Hamilton_2004}.
Based on the vectors $\hat n$ determined from the
$r_l$ statistics (\ref{Eq:rl_statistic}),
an average alignment angle $\hat \theta$ for $l=2$ to 5 is constructed
in \cite{Cresswell_Liddle_Mukherjee_Riazuelo_2006}
and also applied to slab topologies.
The very small average alignment angle $\hat \theta$ obtained
from the TdOH map \cite{Tegmark_deOliveira_Costa_Hamilton_2003}
is unprobable in simply-connected spaces.
It is claimed \cite{Cresswell_Liddle_Mukherjee_Riazuelo_2006}
that a modest increase in the probability for such a
$\hat \theta$ value can be obtained for very thin slab spaces.
However, as mentioned above, the $r_l$ statistics is very sensitive
to noise and, indeed, other full-sky maps possess much larger
values of $\hat \theta$
as it is the case for the foreground-cleaned maps
in the Q, V, and W bands \cite{Bennett_et_al_2003b}
and the ILC map derived from the former, as well as for
the ``Lagrange-ILC'' map of \cite{Eriksen_Banday_Gorski_Lilje_2004}.
Thus it may be unnecessary to require very small values of $\hat \theta$.

There arises the question what large-angle CMB properties
other multi-connected space forms have
with respect to the alignment between quadrupole and octopole, and
with respect to the multipole vector representation.
To that aim we consider in this paper as space manifolds ${\cal M}$
three spherical space forms, the toroidal universes in flat space,
and the Picard topology in hyperbolic space.
The quadrupole-octopole alignment and the multipole vectors
corresponding to $l=2$ and $l=3$ are investigated for
these space forms.

\section{The multi-connected space forms}
\label{The_multi_connected_space_forms}

The most remarkable signature of multi-connected universes
is a suppression of the CMB power spectrum at large angular scales,
in particular a large suppression of the CMB quadrupole and octopole,
and of the temperature two-point correlation function at large angles.
It is exactly this property which neatly explains the low power
at large angles as first observed by COBE \cite{Hinshaw_et_al_1996}
and later substantiated by WMAP \cite{Bennett_et_al_2003}.
For an introduction to ``cosmic topology'',
see \cite{Lachieze-Rey_Luminet_1995,Levin_2002}.

It is claimed in \cite{Luminet_Weeks_Riazuelo_Lehoucq_Uzan_2003}
that the Poincar\'e dodecahedron
(corresponding to the binary icosahedral group $I^\star$)
explains well the WMAP data on large scales.
Since the CMB computations in \cite{Luminet_Weeks_Riazuelo_Lehoucq_Uzan_2003}
are based on very few eigenmodes only,
a comparison to the WMAP data using much more eigenmodes is carried out
in \cite{Aurich_Lustig_Steiner_2004c}.
Besides the Poincar\'e dodecahadron, there are two further good
candidates for multi-connected spherical space forms
\cite{Aurich_Lustig_Steiner_2005a} corresponding to the
binary tetrahedral group $T^\star$ and the
binary octahedral group $O^\star$.
Spherical topologies based on cyclic groups $Z_k$ or on
binary dihedral groups $D_{4m}^\star$ do not lead to a sufficient
suppression on large scales
\cite{Uzan_Riazuelo_Lehoucq_Weeks_2003,Aurich_Lustig_Steiner_2005a}.
Here, we investigate the alignment and multipole vector properties of
the binary tetrahedron ${\cal S}^3/T^\star$,
the binary octahedron ${\cal S}^3/O^\star$, and
the Poincar\'e dodecahedron ${\cal S}^3/I^\star$.

In flat space there are 17 multi-connected space forms
out of which 10 possess a finite volume.
From the latter ones we choose the simplest space form,
i.\,e.\ the hypertorus $T^3$, where opposite faces are identified
without any rotation.
In contrast to space forms of non-flat universes,
the size of the space forms in flat universes is independent of
the curvature, i.\,e.\ of $\Omega_{\hbox{\scriptsize tot}}$
and can thus be chosen freely.
Below, the side lengths $L$ of the hypertori will be given in units of
the Hubble radius $a_0 = \frac{c}{H_0}$.
For the selected cosmological parameters the radius of the
surface of last scattering (SLS) is
$\eta_{\hbox{\scriptsize SLS}} = 3.267$.

The above described space forms are homogeneous.
An example of a non-homogeneous space form is the Picard topology
of hyperbolic space.
This model is discussed in the framework of cosmology in
\cite{Aurich_Lustig_Steiner_Then_2004a,Aurich_Lustig_Steiner_Then_2004b}.
Although its fundamental cell possesses an infinitely long horn,
and is thus non-compact, it has a finite volume.
The shape of the fundamental cell is that of a hyperbolic pyramid with one
cusp at infinity, see \cite{Aurich_Lustig_Steiner_Then_2004a}
for details.
Because of its inhomogeneity, the considered statistics depends on the
position of the observer.
In the following we choose two positions, one ``near'' to the cusp and
the other ``far away'' from it.
For definiteness, choosing the upper half-space as the model of
hyperbolic space with Gaussian curvature $K=-1$,
the two observers $A$ and $B$ are at $(x_1,x_2,x_3) = (0.2, 0.1, 1.6)$
and $(0.2, 0.1, 5.0)$, respectively
\cite{Aurich_Lustig_Steiner_Then_2004a}.

This non-compact space form has the special property that the spectrum
corresponding to the Laplace-Beltrami operator is not purely discrete;
instead its discrete spectrum is imbedded in a continuous spectrum.
The eigenfunctions of the discrete spectrum are the Maa\ss\
cusp forms and those of the continuous spectrum are given by
an Eisenstein series.
The two types of eigenfunctions have different properties and
are discussed separately in the following.

\section{The quadrupole-octopole alignment}

The angular momentum dispersion (\ref{Eq:AMD})
leads for each multipole $l$ to one unit vector $\hat n_l$
as discussed in the Introduction.
The quadrupole-octopole alignment was revealed in
\cite{Tegmark_deOliveira_Costa_Hamilton_2003,%
deOliveira-Costa_Tegmark_Zaldarriaga_Hamilton_2004}
by considering the dot product
\begin{equation}
\label{Eq:Alignment_A23}
A_{23} \; := \; | \, \hat n_2 \cdot \hat n_3 \, |
\hspace{10pt} .
\end{equation}
For the TdOH sky map the surprisingly high value of $A_{23} \simeq 0.9849$
\cite{deOliveira-Costa_Tegmark_Zaldarriaga_Hamilton_2004} was found
corresponding to an angular separation between $\hat n_2$ and $\hat n_3$
of only $10^\circ$.
This result is essentially confirmed using the WMAP 3yr measurements
independent of the used mask to eliminate galactic foregrounds
\cite{deOliveira-Costa_Tegmark_2006}.
Under the assumption that the CMB is an isotropic random field,
all multipoles are statistically independent and all directions
of the $\hat n_l$'s are equally probable.
In this case the values of $A_{23}$ are {\it uniformly} distributed on the
interval $[0,1]$.
A value greater than $A_{23} = 0.9849$ is thus obtained only for
$1.51\%$ of isotropic CMB realizations.

The root of the quadrupole-octopole alignment is the unusual behaviour
of the $a_{lm}$'s for $l=2$ and $l=3$.
They have the property that there exists a coordinate system
in which all values of $|a_{lm}|$ are very small except
$|a_{ll}|$ and $|a_{l0}|$
(see Table III in \cite{deOliveira-Costa_Tegmark_Zaldarriaga_Hamilton_2004}).
Since the value of $|a_{l0}|$ is irrelevant for the
angular momentum dispersion (\ref{Eq:AMD}),
the alignment points to models of the universe
having small values of $|a_{lm}|/|a_{ll}|$ for $m=1,\dots,l-1$.
Two conditions are necessary for this to be the case:
\begin{itemize}
\item[i)]
The lowest eigenfunctions of the multi-connected space ${\cal M}$ should
possess spherical expansion coefficients $a_{lm}(n,i)$
which satisfy the topological alignment condition
$$
\frac{|a_{lm}(n,i)|}{|a_{ll}(n,i)|} \; \ll \; 1
\hspace{10pt} \hbox{ for } \hspace{10pt}
m=1,\dots,l-1
\hspace{10pt} \hbox{ and } \hspace{10pt}
l = 2,3
\hspace{10pt} .
$$
Here $n$ counts the eigenvalues and $i$ is the degeneracy index
$i=1,\dots, r^{\cal M}(n)$, where $r^{\cal M}(n)$ is the multiplicity
of the $n$th eigenvalue.
\item[ii)]
The transfer function $T_l(k)$ \cite{Hu_1995} should emphasize those
lowest eigenfunctions with wave-number $k=k_n$
which have the property i) and suppress all others.
\end{itemize}
One has to enforce condition i) for the lowest eigenfunctions
because they give the most important contribution
to the CMB fluctuations at large scales and
because the separation between consecutive eigenmodes is most
pronounced in the lower part of the spectrum.
Property i) is so unusual that only a few eigenmodes will possess it.
At the higher part of the spectrum the eigenmodes are lying so close
together that the transfer functions and the random expansion coefficients
will statistically destroy the alignment.

Whereas property i) is determined by the multi-connected space form,
property ii) is determined by the cosmological parameters.
In the flat case one has, in addition, the freedom to scale the size
of the space form instead of the cosmological parameters.
Thus it does not suffice to have a manifold with alignment,
one also needs cosmological parameters leading to a
transfer function which enables the survival of the alignment
in the CMB fluctuations.

\subsection{The flat hypertorus}

\begin{figure}[htb]
\begin{center}
\vspace*{-80pt}
\hspace*{-110pt}\begin{minipage}{14cm}
\begin{minipage}{6cm}
\includegraphics[width=9.0cm]{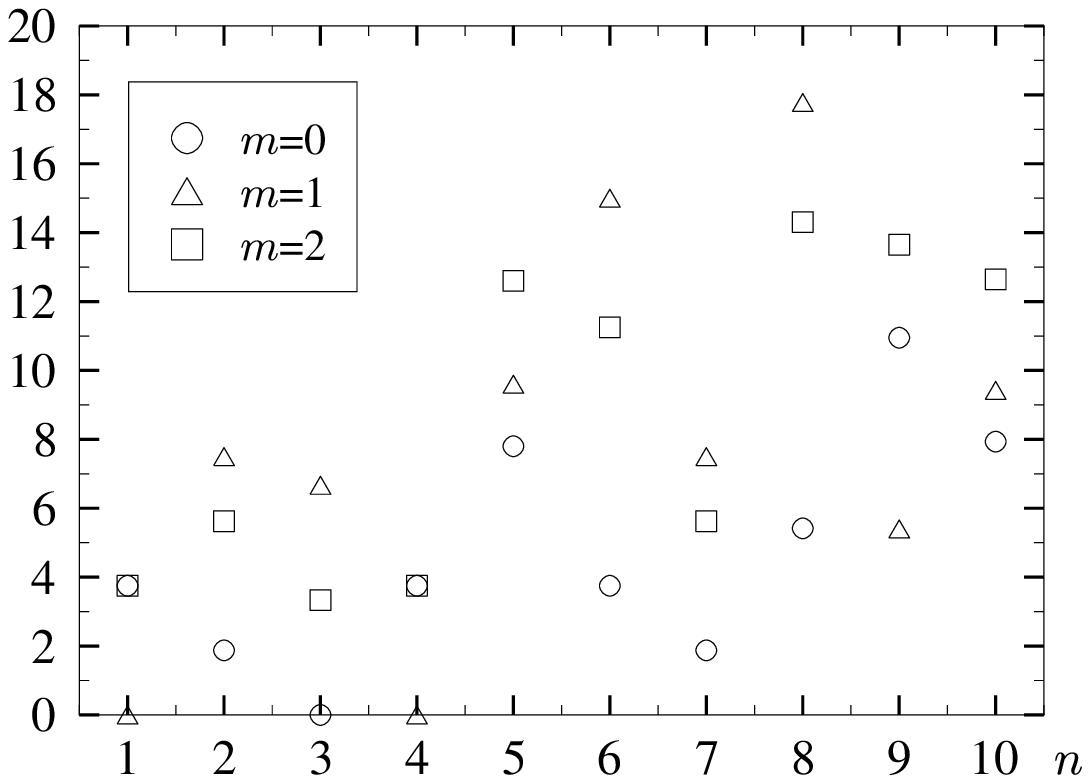}
\put(-40,155){(a)}
\end{minipage}
\begin{minipage}{6cm}
\hspace*{50pt}\includegraphics[width=9.0cm]{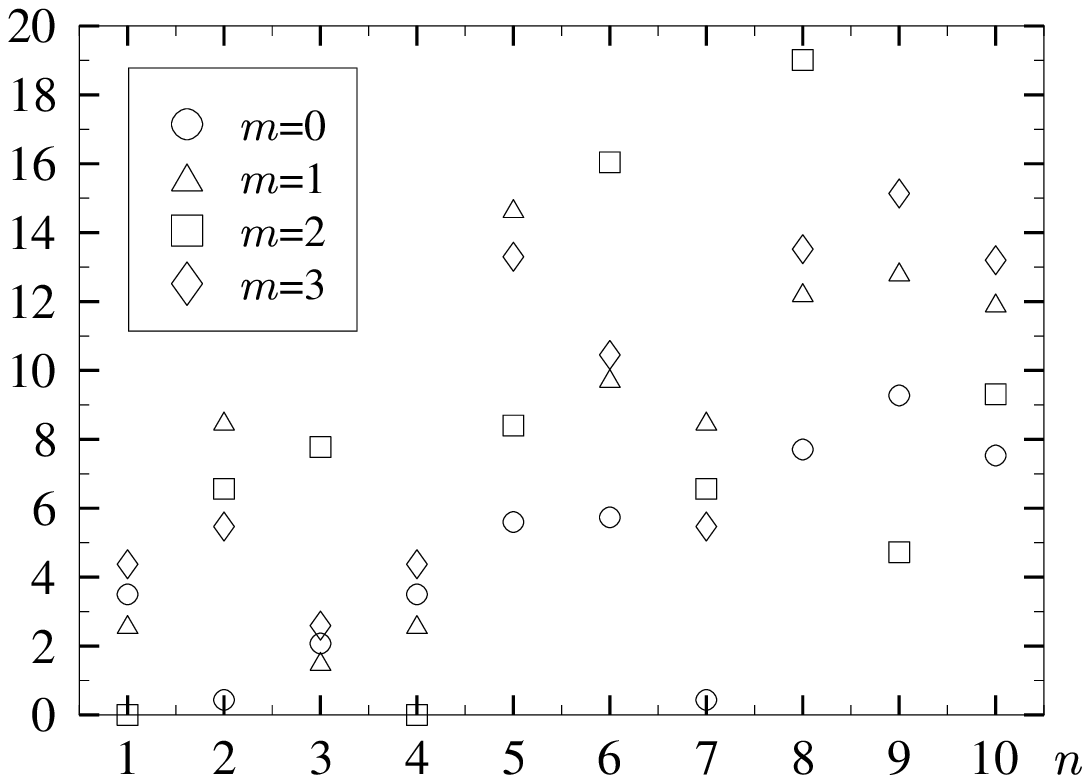}
\put(-40,155){(b)}
\end{minipage}
\end{minipage}
\end{center}
\vspace*{-40pt}
\caption{\label{Fig:b_lm_torus}
The topological alignment measure (\ref{Eq:b_lm})
is shown for the hypertorus.
Panel (a) shows the values of $b_{lm}(n)$ for $l=2$
for the first ten eigenvalues.
Panel (b) displays the analogous quantities for $l=3$.
}
\end{figure}

Let us demonstrate this dependence on the transfer function
in the case of the flat hypertorus,
where all three sides are chosen to be equal.
The fundamental cell is thus a cube and the lowest eigenvalue
is sixfold degenerate.
All six eigenfunctions are surprisingly close to our property i).
Let us define the {\it topological alignment measure} $b_{lm}(n)$
as the following sum over all eigenfunctions belonging to the
$n$th-eigenvalue $k_n^2$
\begin{equation}
\label{Eq:b_lm}
b_{lm}(n) :=
\sum_{i=1}^{r^{\cal M}(n)} \big|a_{lm}(n,i)\big|^2 \; = \;
\sum_{i=1}^{r^{\cal M}(n)}
\left| Y_{lm}\left(\frac{\vec k_i}{|\vec k_i|}\right) \right|^2
\hspace{10pt} ,
\end{equation}
where $a_{lm}(n,i)$ is the spherical expansion coefficient
of the eigenfunction belonging to the wavevectors $\vec k_i$
with $i=1,\dots,r^{\cal M}(n)$.
To obtain from (\ref{Eq:b_lm}) a unique quantity,
one has to choose the value of equation (\ref{Eq:b_lm}) for that
rotated coordinate system for which 
the property i) is optimally satisfied.
Figure \ref{Fig:b_lm_torus} shows that the hypertorus with
equal side lengths has ground state eigenfunctions with are nearly optimal.
For $l=2$ all six eigenfunctions possess $a_{21}(1,i)=0$ and thus,
$b_{21}(1)=0$ is obtained.
In the case $l=3$ the measure $b_{32}(1)$ vanishes,
but $b_{31}(1)$ is not zero.
Would the latter also be zero, the first six eigenfunctions belonging
to the first eigenvalue would all have the optimal topological alignment.
Figure \ref{Fig:b_lm_torus} reveals that the next eigenvalues
do not possess such a nice alignment behaviour,
although there are some  special cases like $b_{21}(4)$ or
$b_{32}(4)$, which vanish as in the case $n=1$.


\begin{figure}[htb]
\begin{center}
\vspace*{-80pt}
\hspace*{-110pt}\begin{minipage}{14cm}
\begin{minipage}{6cm}
\includegraphics[width=9.0cm]{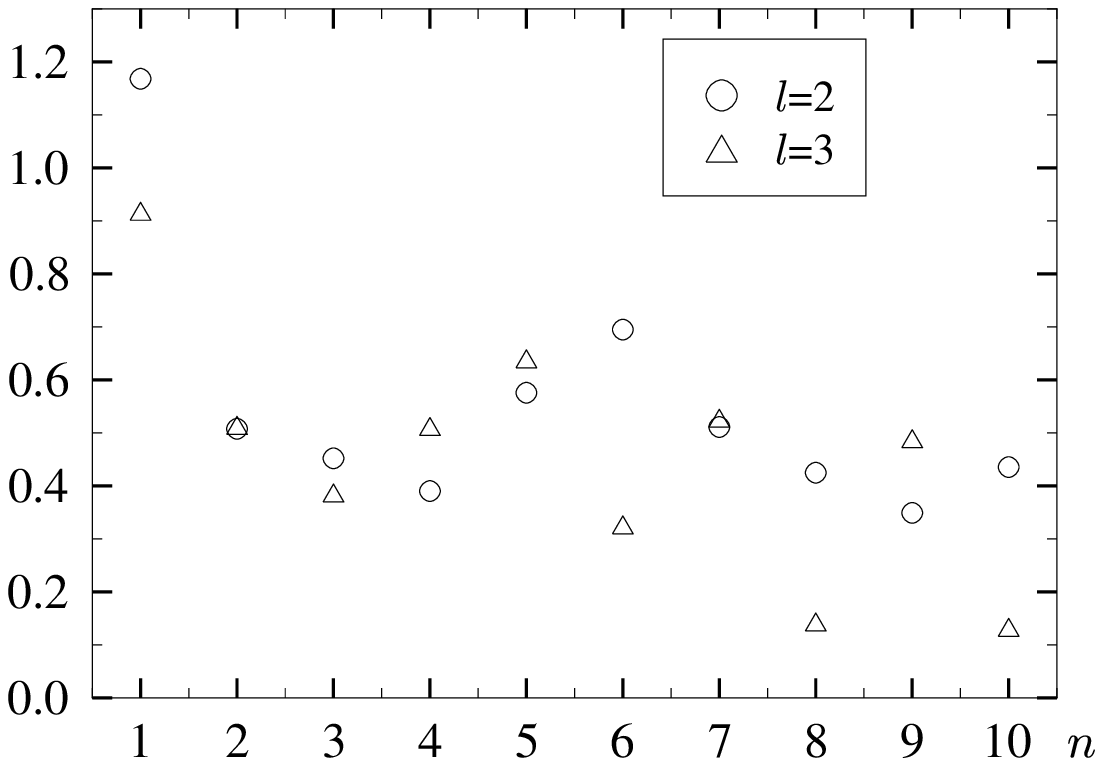}
\put(-35,155){(a)}
\end{minipage}
\begin{minipage}{6cm}
\hspace*{50pt}\includegraphics[width=9.0cm]{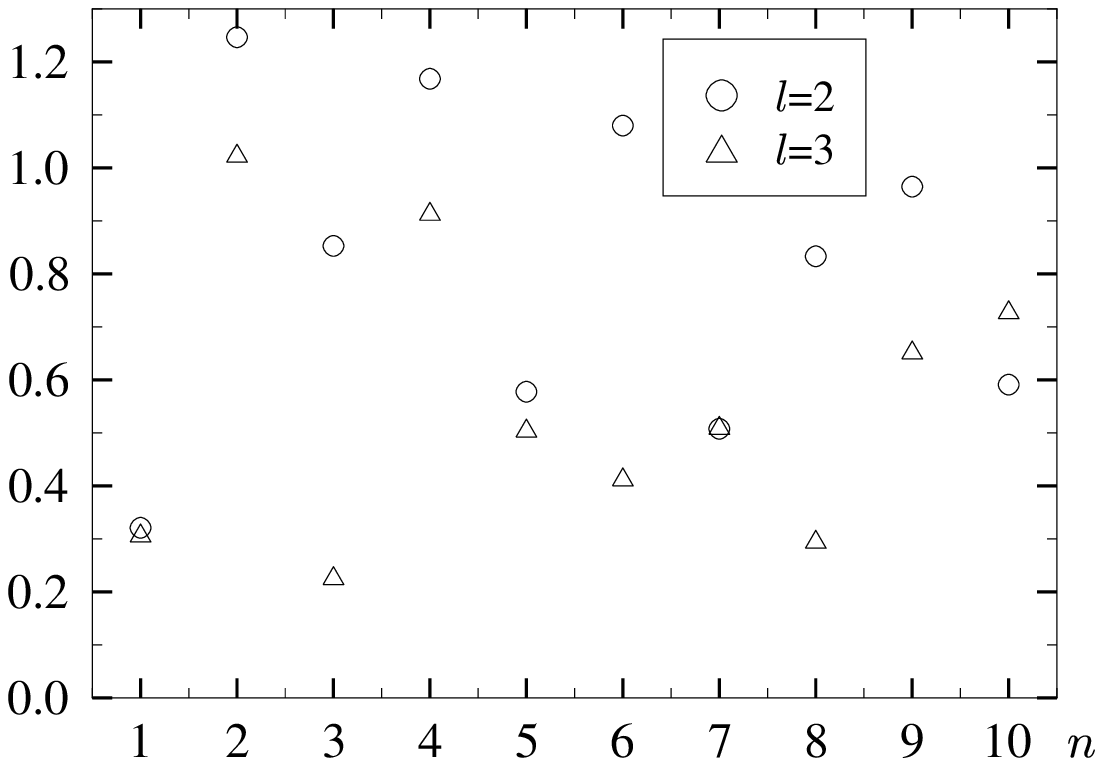}
\put(-35,155){(b)}
\end{minipage}
\end{minipage}
\end{center}
\vspace*{-40pt}
\caption{\label{Fig:transfer_torus_L_1_2}
The transfer function $T_l(k_n)$ is shown for a hypertorus
with side length $L=1.0$ (a) and with $L=2.0$ (b).
The abscissae give the number of the eigenvalue $n$
instead of the eigenvalue $k_n$.
}
\end{figure}

This is the point where the transfer function $T_l(k)$ becomes important.
According to our property ii) one needs a transfer function,
which emphasizes exactly those eigenvalues having a maximal
topological alignment.
In our example a transfer function $T_l(k)$ which is large
for $l=2$ and $l=3$ at $n=1$ and $n=4$ and small at $n=2, 3, 5$
would produce more aligned CMB sky maps than a model with
a simply-connected space form.
In figure \ref{Fig:transfer_torus_L_1_2}
the transfer function $T_l(k)$ is shown for the cosmological parameters
$\Omega_{\hbox{\scriptsize bar}} = 0.046$,
$\Omega_{\hbox{\scriptsize cdm}} = 0.234$,
$\Omega_\Lambda=0.72$, and $h=0.7$.
Here, reionization is omitted having only a modest influence on
such large scales like $l=2$ and $l=3$,
and a standard thermal history of the neutrinos is assumed.
The side length $L$ of the cube determines the scaling of the
eigenvalue spectrum.
Panel (a) of figure \ref{Fig:transfer_torus_L_1_2} shows the
transfer function $T_l(k)$ for a hypertorus with $L=1.0$,
whereas in panel (b) $L=2.0$ is chosen.
One observes in figure \ref{Fig:transfer_torus_L_1_2}
that for $L=1.0$ the first eigenvalue is emphasized by $T_l(k_n)$
for $l=2$ and $l=3$.
For $L=2.0$ the opposite behaviour is revealed,
i.\,e.\ the contribution of the first six eigenfunctions
belonging to the first eigenvalue is suppressed.
Thus, one expects varying deviations from the
uniform distribution for $A_{23}$,
eq.\,(\ref{Eq:Alignment_A23}), depending on the side length $L$.


\begin{figure}[htb]
\begin{center}
\vspace*{-80pt}
\hspace*{-80pt}\begin{minipage}{14cm}
\begin{minipage}{6cm}
\includegraphics[width=9.0cm]{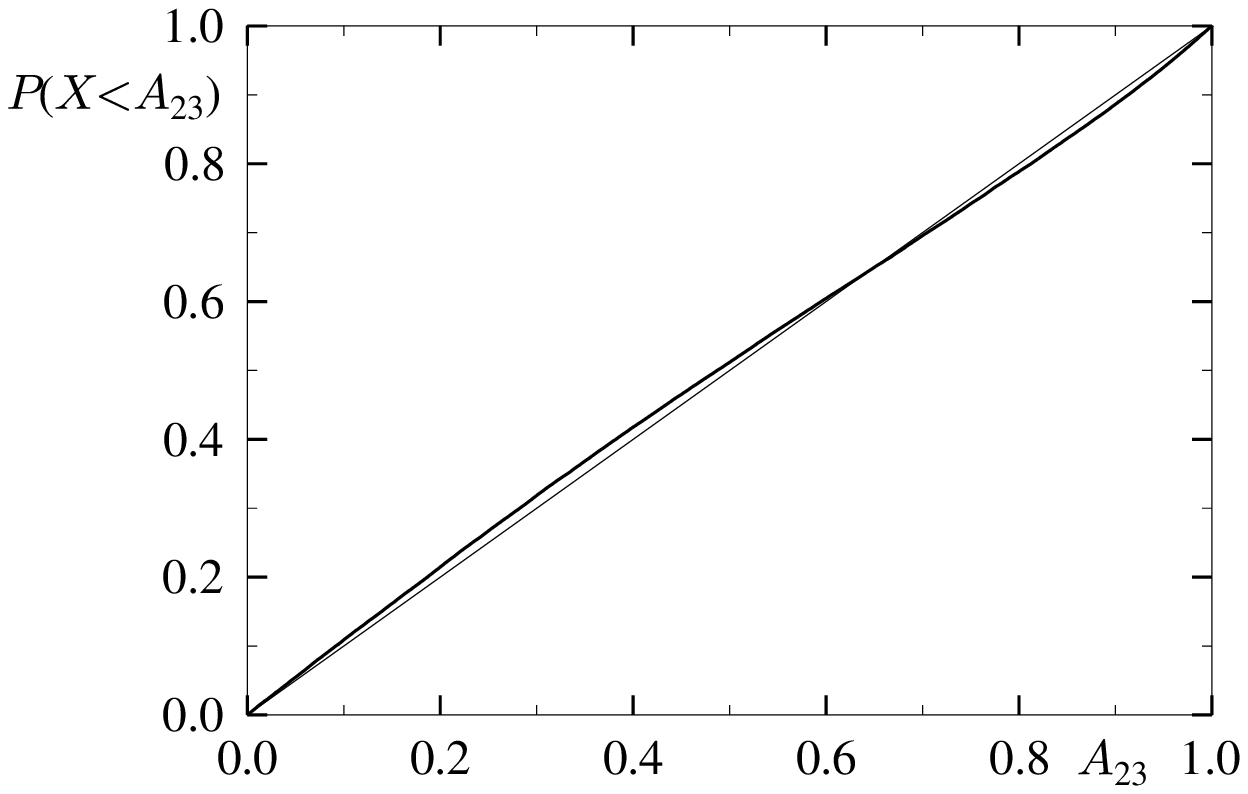}
\put(-180,155){(a)}
\end{minipage}
\begin{minipage}{6cm}
\hspace*{50pt}\includegraphics[width=9.0cm]{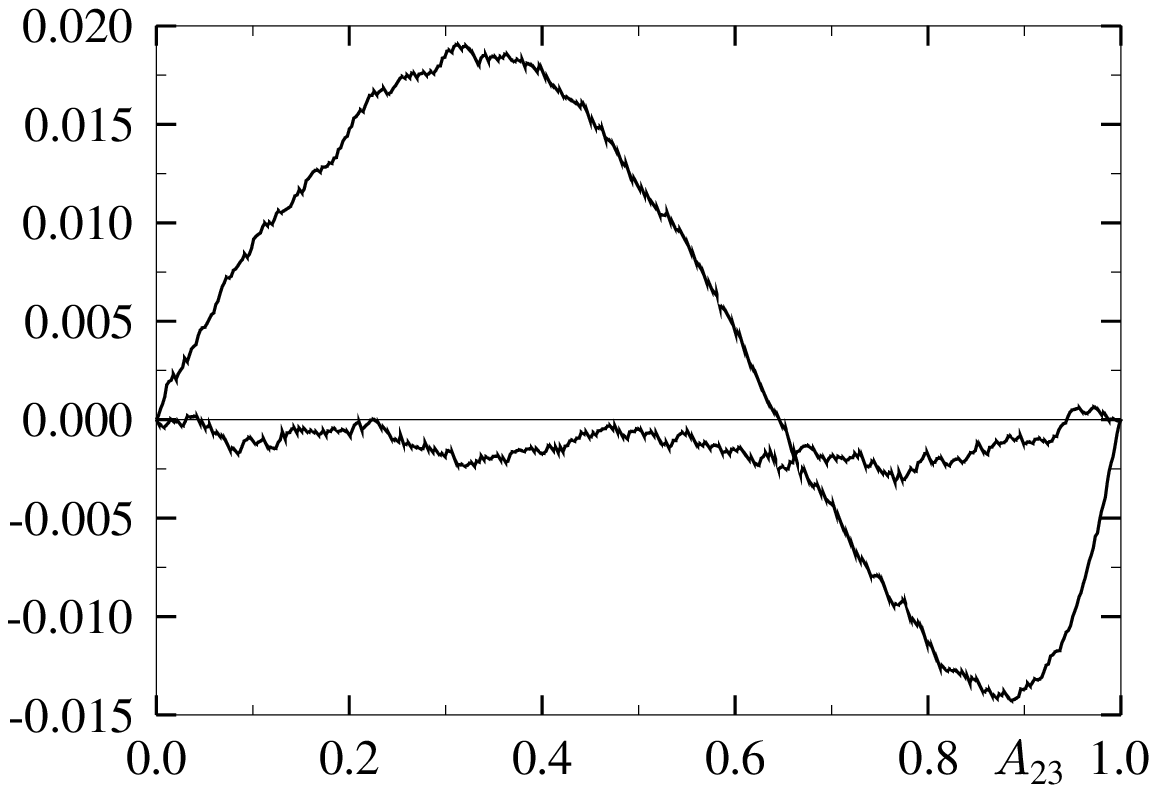}
\put(-175,155){(b)}
\end{minipage}
\end{minipage}
\end{center}
\vspace*{-40pt}
\caption{\label{Fig:A23_cumu_torus}
The cumulative distribution $P(X<A_{23})$ is shown for the hypertorus
with side length $L=1.0$ together with the uniform distribution
(straight line) in panel (a).
Since the curve for $L=2.0$ would be indistinguishable
from the uniform distribution,
panel (b) shows the difference to the uniform distribution
for the torus with side length $L=1.0$ (large deviations from the zero line)
and with side length $L=2.0$ (small deviations), respectively.
}
\end{figure}

Now we discuss the value distribution of $A_{23}$
for the hypertorus in dependence on the side length $L$.
For each given cube specified by $L$ we generate 100\,000 sky realizations
with a wave number cut-off $k = 60$ in units of the Hubble length.
This allows an accurate determination of
the maximum angular momentum dispersion vectors
$\hat n_2$ and $\hat n_3$ in order to compute $A_{23}$.
The results for the flat hypertorus are shown in
figure \ref{Fig:A23_cumu_torus}.
In panel (a) the cumulative distribution is shown for
100\,000 realizations for a torus with side length $L=1.0$
together with the uniform distribution.
The cumulative distribution for $L=2.0$ would be indistinguishable
from the uniform distribution in this figure.
Thus panel (b) displays the difference to the uniform distributions
for $L=1.0$ and $L=2.0$.
As expected from the above discussion,
the case $L=1.0$ where the transfer function $T_l(k_n)$ enhances the
contribution of the first eigenvalue displays large deviations from
the uniform distribution,
whereas in the other case, the obtained distribution is nearly uniform.
This demonstrates that one not only needs a space form
with ``aligned'' eigenfunctions but also suitable cosmological parameters.


\begin{figure}[htb]
\begin{center}
\vspace*{-80pt}
\hspace*{-80pt}\begin{minipage}{14cm}
\begin{minipage}{6cm}
\includegraphics[width=9.0cm]{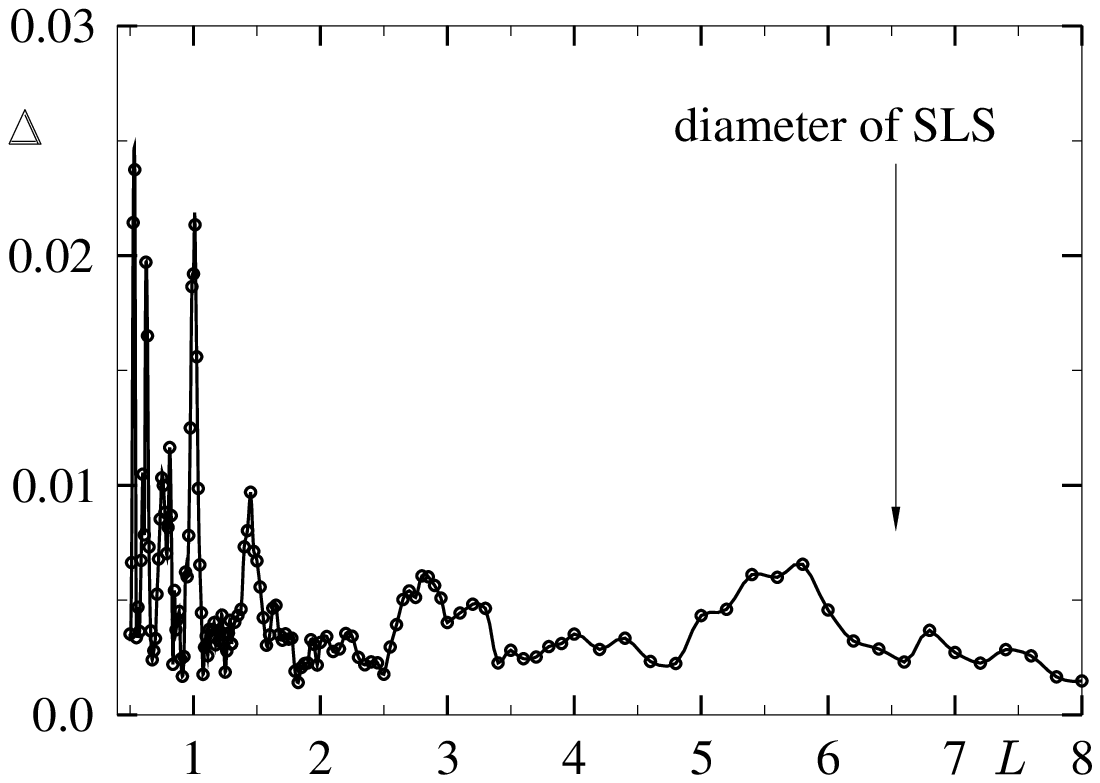}
\put(-170,155){(a)}
\end{minipage}
\begin{minipage}{6cm}
\hspace*{50pt}\includegraphics[width=9.0cm]{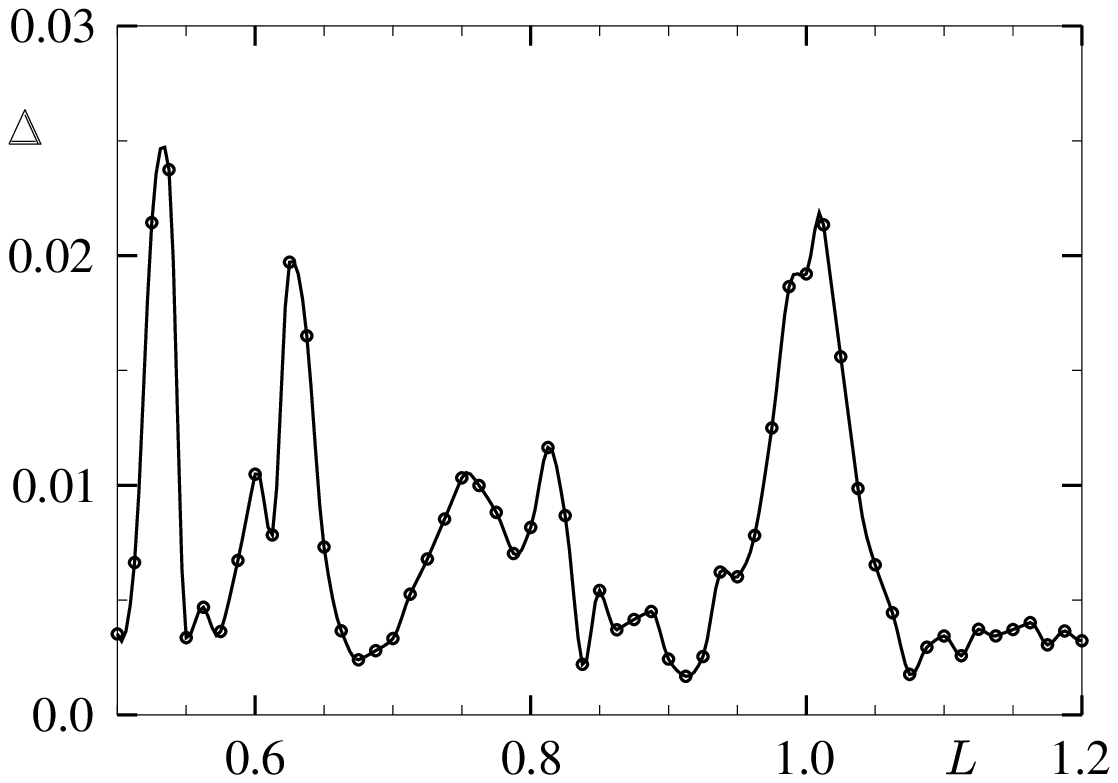}
\put(-40,155){(b)}
\end{minipage}
\end{minipage}
\end{center}
\vspace*{-40pt}
\caption{\label{Fig:A23_deviation_torus}
The maximal deviation $\Delta$ of the cumulative distribution $P(X<A_{23})$
from the uniform distribution is shown in dependence of the side length $L$
of the hypertori.
The diameter of the SLS at $2\eta_{\hbox{\scriptsize SLS}} = 6.534$
is indicated.
Models with a side length $L$ below this value are completely within the SLS.
}
\end{figure}

\begin{figure}[htb]
\begin{center}
\vspace*{-80pt}
\hspace*{-80pt}\begin{minipage}{14cm}
\begin{minipage}{6cm}
\includegraphics[width=9.0cm]{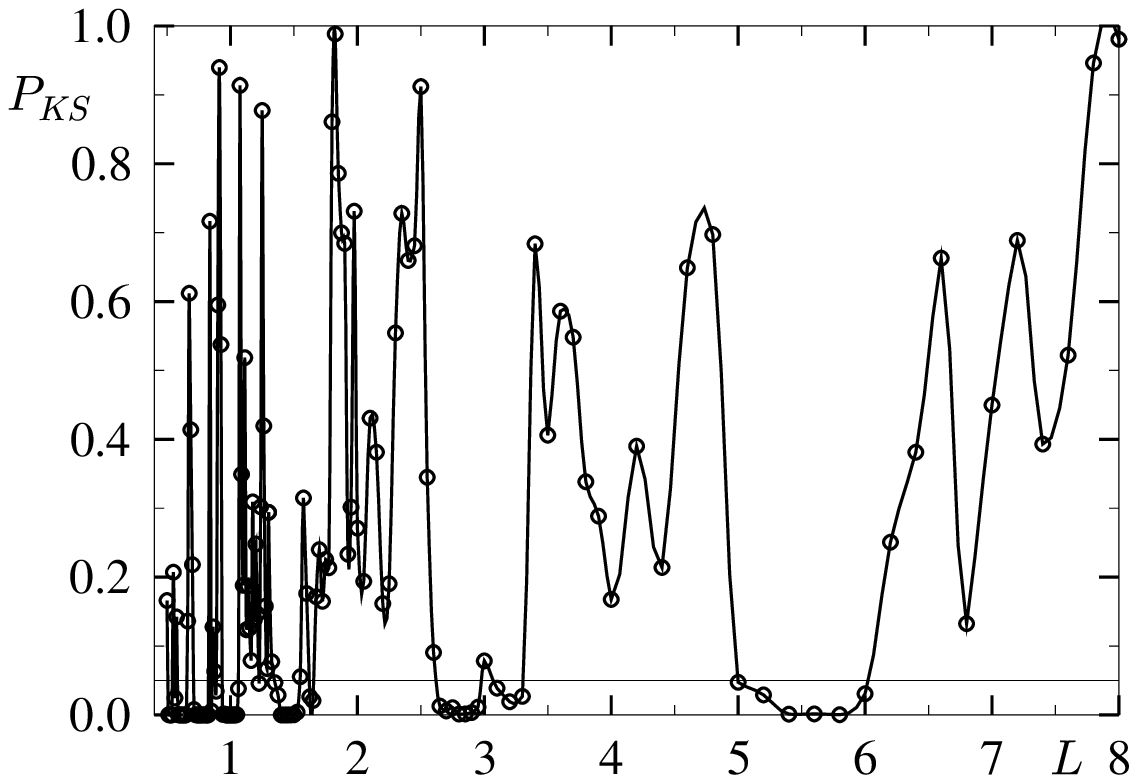}
\put(-35,155){(a)}
\end{minipage}
\begin{minipage}{6cm}
\hspace*{50pt}\includegraphics[width=9.0cm]{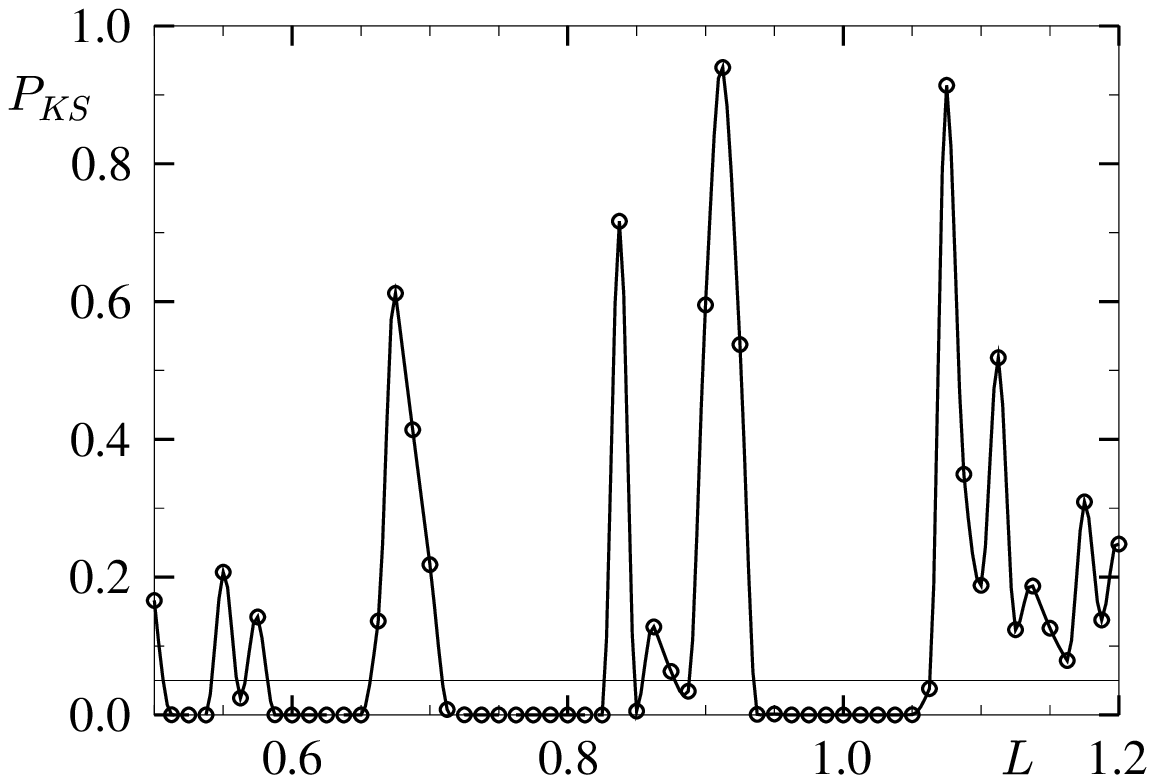}
\put(-35,155){(b)}
\end{minipage}
\end{minipage}
\end{center}
\vspace*{-40pt}
\caption{\label{Fig:Torus_A23_Kolmo_Smirnov}
The Kolmogorov-Smirnov test is applied to the distribution of the
values of $A_{23}$ compared to the uniform distribution for
hypertori with side lengths $L$ covering the interval $L\in[0.5,8]$.
}
\end{figure}

To address the question whether the deviations are significant,
the Kolmogorov-Smirnov test is applied to the distribution of $A_{23}$
with respect to the uniform distribution.
This test gives the probability $P_{KS}$
that the maximal deviation $\Delta$ from an assumed distribution is in
accordance with Gaussian fluctuations given a finite set of $N$ data points.
The maximal deviation $\Delta$ is shown for a large sequence of side lengths
$L$ in figure \ref{Fig:A23_deviation_torus}.
For each side length $L$ denoted by a small circle,
$N=100\,000$ simulations have been carried out.
One recognises the large deviation at the side length $L=1.0$
as already discussed above as well as only a small deviation
in the case $L=2.0$.
Once again, the importance of the transfer function $T_l(k_n)$ is obvious
since all eigenfunctions possess the same topological
alignment $b_{lm}(n)$ as seen in figure \ref{Fig:b_lm_torus}.
In figure \ref{Fig:Torus_A23_Kolmo_Smirnov} the
Kolmogorov-Smirnov probability $P_{KS}$ is displayed with respect to
the $N=100\,000$ simulations.
The horizontal line lies at a probability of 5\%.
Hypertori below this significance level can be considered as having
a too large maximal deviation $\Delta$ to be compatible with
statistical fluctuations.
These are the hypertori whose values of $A_{23}$ are not uniformly distributed.


\begin{figure}[htb]
\begin{center}
\vspace*{-80pt}
\hspace*{-80pt}\begin{minipage}{14cm}
\begin{minipage}{6cm}
\includegraphics[width=9.0cm]{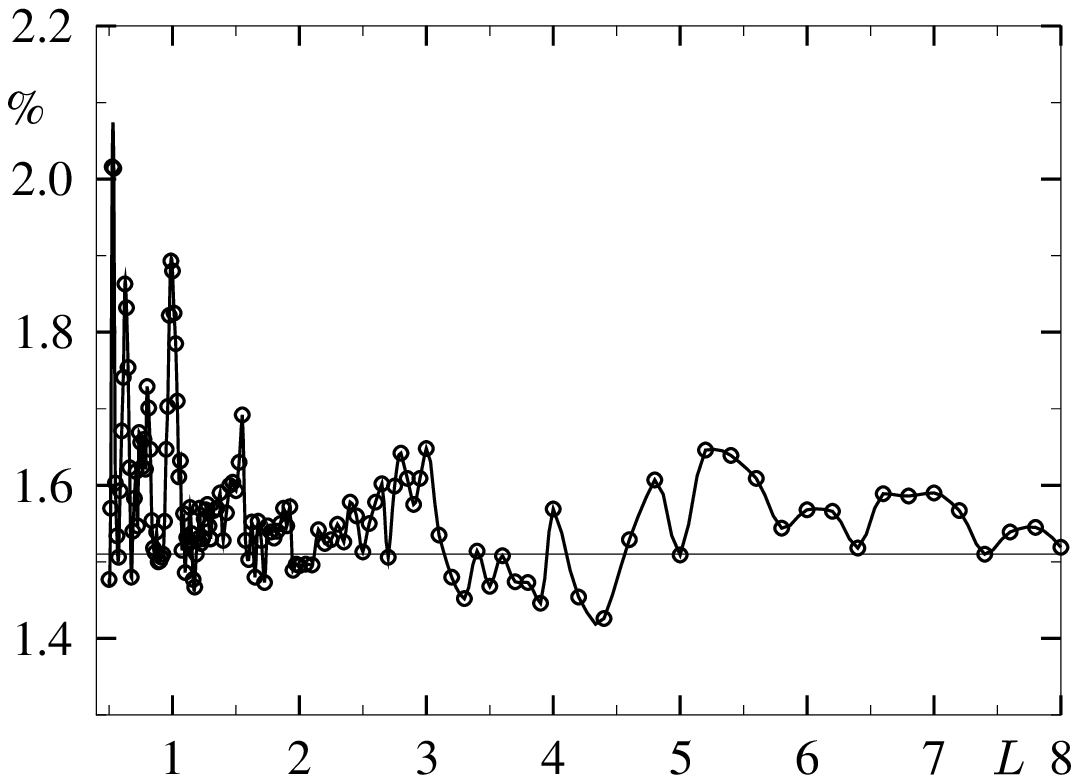}
\put(-35,155){(a)}
\end{minipage}
\begin{minipage}{6cm}
\hspace*{50pt}\includegraphics[width=9.0cm]{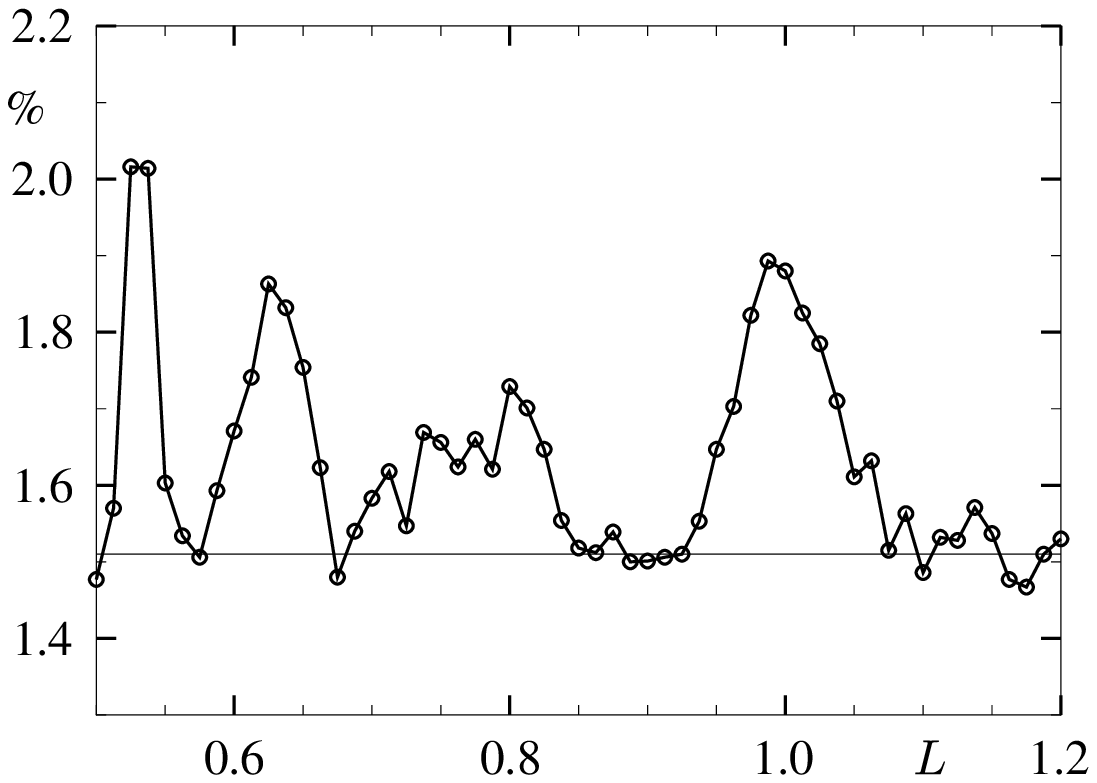}
\put(-35,155){(b)}
\end{minipage}
\end{minipage}
\end{center}
\vspace*{-40pt}
\caption{\label{Fig:Torus_A23_Prozent_0.9849}
The percentile of the models having a larger value of $A_{23}$
than the observed value of $A_{23}=0.9849$ is shown.
For a uniform distribution one expects a percentile of 1.51\%
which is indicated by the horizontal line.
}
\end{figure}

The observed value of $A_{23}$ very close to one can hardly be
explained by the flat hypertori.
A value greater than the observed one of $A_{23} = 0.9849$ is obtained
only for $1.51\%$ of isotropic CMB realizations.
In figure \ref{Fig:Torus_A23_Prozent_0.9849} this value is compared with
our sequence of tori.
For those tori where the Kolmogorov-Smirnov test signals significant
deviation from the uniform expectation,
one observes indeed a higher percentile up to roughly 2\% of models
having larger values than $A_{23} = 0.9849$.
One gets for these model an increased probability for the alignment
but it is a matter of personal judgement
whether one considers a 1:50 probability as a serious problem.


\begin{figure}[htb]
\begin{center}
\vspace*{-70pt}
\hspace*{-90pt}\begin{minipage}{14cm}
\begin{minipage}{6cm}
\includegraphics[width=9.0cm]{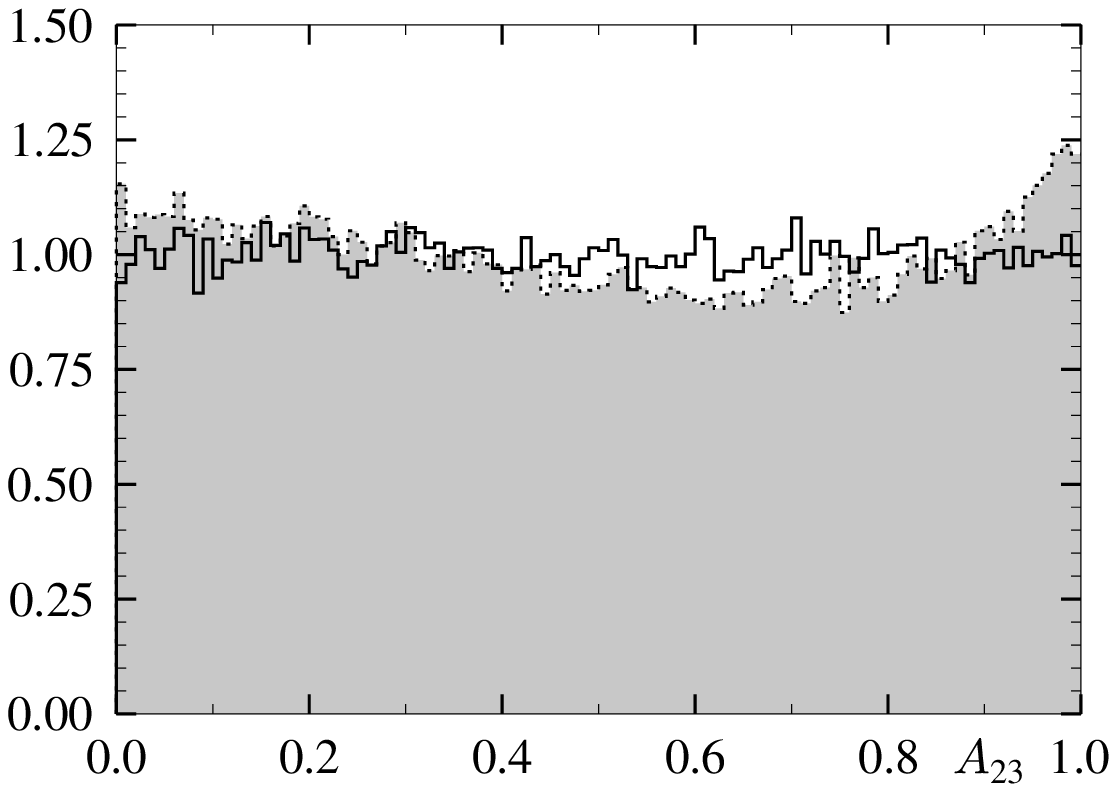}
\put(-170,155){(a)}
\end{minipage}
\begin{minipage}{6cm}
\hspace*{60pt}\includegraphics[width=9.0cm]{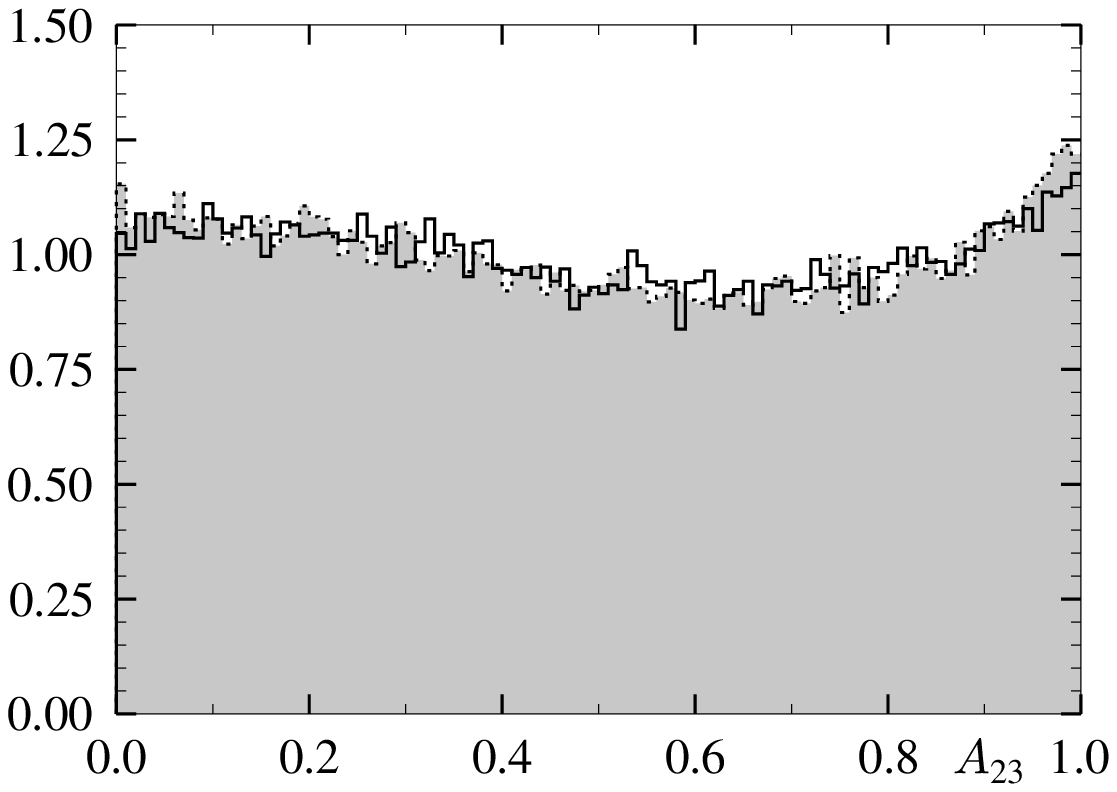}
\put(-175,155){(b)}
\end{minipage}
\end{minipage}
\hspace*{-90pt}\begin{minipage}{14cm}
\vspace*{-90pt}\begin{minipage}{6cm}
\includegraphics[width=9.0cm]{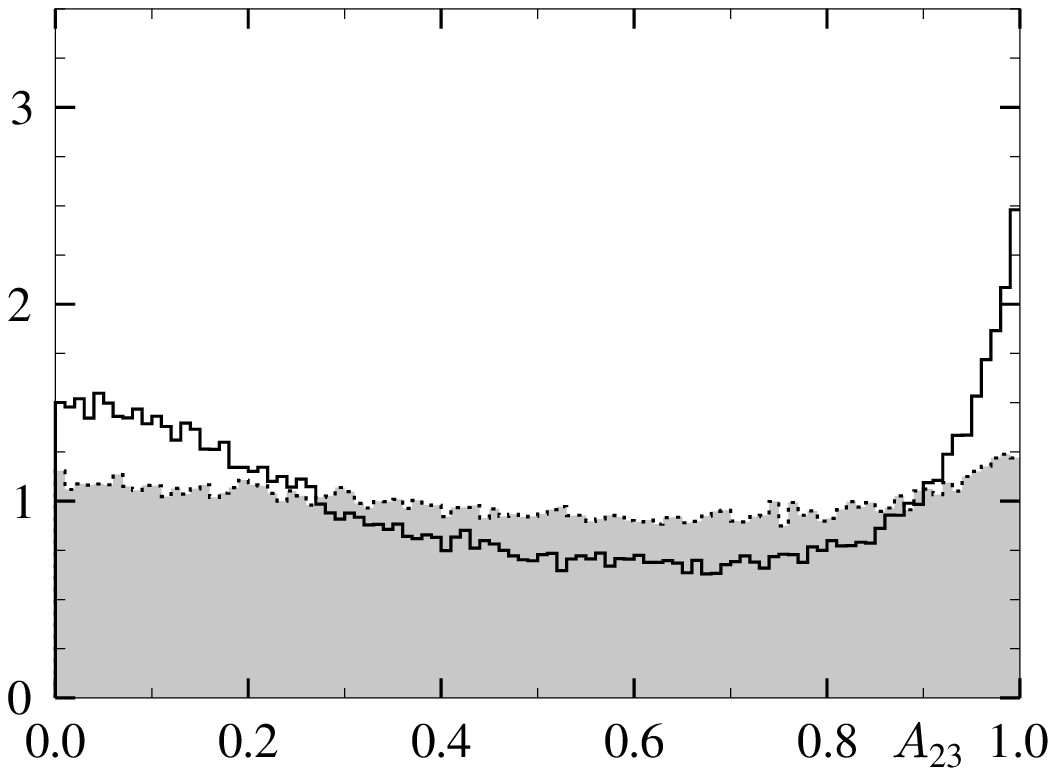}
\put(-180,155){(c)}
\end{minipage}
\begin{minipage}{6cm}
\hspace*{60pt}\includegraphics[width=9.0cm]{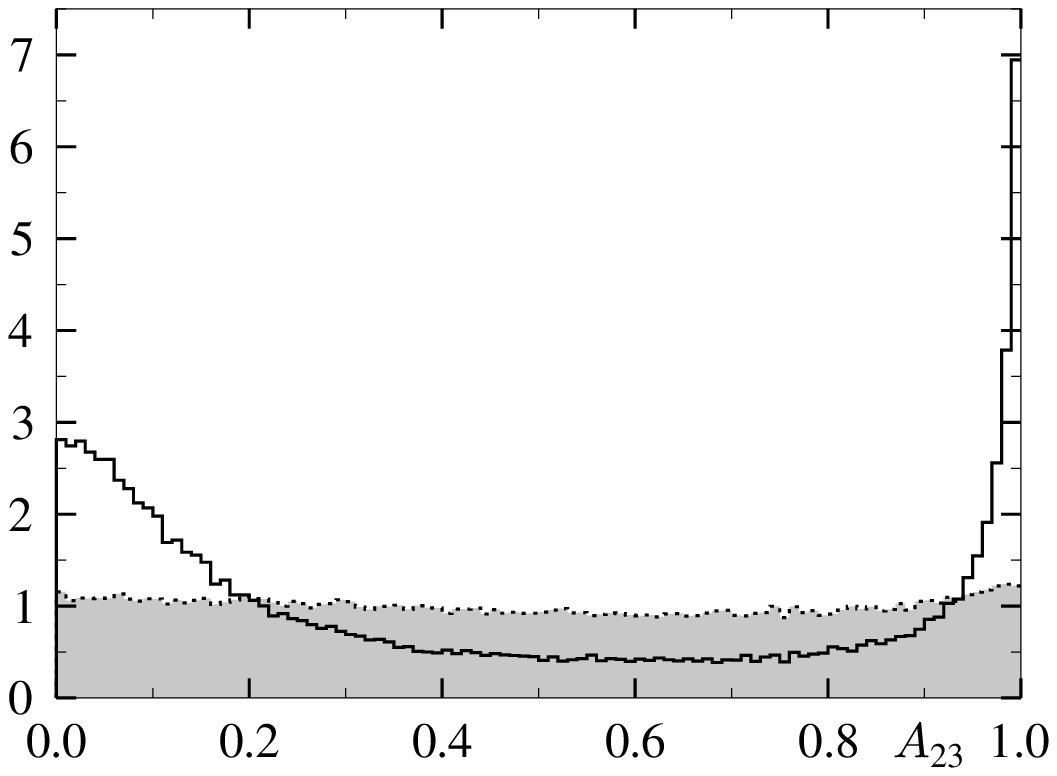}
\put(-175,155){(d)}
\end{minipage}
\end{minipage}
\end{center}
\vspace*{-40pt}
\caption{\label{Fig:modified_alm_trans}
The distribution of $A_{23}$ for the hypertorus with side length $L=1.0$
is shown as a dotted curve in all four panels.
As a full curve modified computations are shown:
in panel (a) the contribution of the first eigenvalue $n=1$ is omitted,
in panel (b) the expansion coefficient $a_{31}$ is multiplied by 0.001
for the eigenfunctions belonging to the first eigenvalue,
and in addition to this multiplication, in panels (c) and (d)
the contribution of the eigenvalues $n=2,..,10$
and  $n=2,..,30$ are omitted, respectively.
}
\end{figure}

Let us now demonstrate how difficult it is
to obtain such a strong alignment as it is observed in the CMB sky.
To that aim we return to the cube with side length $L=1.0$
which already possesses a statistically significant deviation
from the uniform distribution.
We now artificially alter some expansion coefficients to increase
the topological alignment or alter the transfer function to enhance the
contribution of the first eigenvalue.
At first we show that the deviation is indeed due to the contribution
belonging to the first eigenvalue.
In figure \ref{Fig:modified_alm_trans}a
the distribution of $A_{23}$ is compared with one where the contribution
of the six eigenfunctions belonging to the first eigenvalue
is artificially omitted.
This omission leads indeed to a equidistribution as seen
in figure \ref{Fig:modified_alm_trans}a.
Whereas 1.880\% of the hypertorus models with $L=1.0$ have a
larger value of $A_{23}$ than the observed value of $A_{23}=0.9849$,
there are only 1.518\% of models with the omitted ground state.
In figure \ref{Fig:modified_alm_trans}b
we alter the spherical expansion coefficients $a_{31}$
by multiplying them by 0.001 for $n=1$.
This leads for the first eigenvalue to a nearly perfect
topological alignment.
However, this does not change the histogram significantly
showing that the next eigenvalues are important too and
destroy the alignment provided by the first eigenvalue.
The percentile of such models with $A_{23}>0.9849$ is 1.770\%.
In the next step we multiply again $a_{31}$ by 0.001 for $n=1$,
but, in addition, we omit the contributions of the second to the
tenth eigenvalue producing artificially a large gap in the
eigenvalue spectrum.
This leads to the distribution shown in figure \ref{Fig:modified_alm_trans}c
possessing a large ``alignment peak'' at $A_{23}=1$.
Further increasing the gap artificially makes this peak more pronounced
as shown in \ref{Fig:modified_alm_trans}d
where the contributions of the eigenvalues from $n=2$ to $n=30$
are omitted.
The corresponding percentiles for $A_{23}>0.9849$ are 3.627\% and
9.103\%, respectively.


In order to find a space form possessing a strong alignment $A_{23}$,
two requirements are important.
On the one hand a space form is needed
where the eigenfunctions of the first eigenvalue possess a strong
topological alignment, property i),
and on the other hand, the gap between the first eigenvalue and
the second should be as large as possible
such that the required property ii) can lead to a survival of
the topological alignment of the first eigenfunctions.

Slab topologies are considered with respect to the
average alignment angle $\hat \theta$
in \cite{Cresswell_Liddle_Mukherjee_Riazuelo_2006}.
There it is found that the average alignment angle $\hat \theta$
deviates from the uniform expectation the more the thinner
the fundamental cells are.
It is an interesting question whether this behaviour is really
monotone or whether a higher resolution with respect to the
thickness of the slabs also reveals a fluctuating behaviour
dictated by the transfer function.

\subsection{Three models with positive spatial curvature}

Let us now discuss the alignment properties of the three
spherical space forms ${\cal S}^3/T^\star$, ${\cal S}^3/O^\star$, and
${\cal S}^3/I^\star$ \cite{Aurich_Lustig_Steiner_2005a}.
Because it requires a large numerical effort to simulate and analyse
a huge number of sky maps,
we do not analyse the parameter space so extensively as in the
case of the hypertorus.
Instead we choose the three different sets of cosmological parameters
for which these three models give a satisfactory description
of the large-scale CMB anisotropy \cite{Aurich_Lustig_Steiner_2005a}.
The cosmological parameters are
$\Omega_{\hbox{\scriptsize bar}} = 0.046$,
$\Omega_{\hbox{\scriptsize cdm}} = 0.234$, and $h=0.7$.
The dark energy contribution varies as $\Omega_\Lambda=0.738$,
$\Omega_\Lambda=0.758$, and $\Omega_\Lambda=0.785$,
which leads to 
$\Omega_{\hbox{\scriptsize tot}} = 1.018$,
$\Omega_{\hbox{\scriptsize tot}} = 1.038$, and
$\Omega_{\hbox{\scriptsize tot}} = 1.065$,
respectively.
These three different values of $\Omega_{\hbox{\scriptsize tot}}$
lead to different sizes of the fundamental cells.
The cells are the larger the closer $\Omega_{\hbox{\scriptsize tot}}$
is to one.
E.\,g.\ for the Poincar\'e dodecahedron ${\cal S}^3/I^\star$,
the surface of last scattering fits completely inside the fundamental cell
below $\Omega_{\hbox{\scriptsize tot}}\simeq 1.01$.
In that case no circles-in-the-sky signature can be observed.

\begin{table*}
\centering
\begin{minipage}{140mm}
\caption{\label{Tab:A23_Spherical}
The maximal deviation $\Delta$ from the uniform distribution and
the corresponding Kolmogorov-Smirnov probability $P_{KS}$
are shown for three spherical spaces for
three sets of cosmological parameters.
The values are based on 100\,000 CMB sky map realizations for each model.
}
\vspace*{3mm}\hspace*{72pt}\begin{tabular}{|c|c|c|c|}
\hline
spherical model & $\Delta$ & $P_{KS}$ & $P(A_{23} > 0.9849)$\\
\hline
${\cal S}^3/T^\star$, $\Omega_{\hbox{\scriptsize tot}} = 1.018$ &
0.00717 & 0.0\% & 1.481\% \\
\hline
${\cal S}^3/T^\star$, $\Omega_{\hbox{\scriptsize tot}} = 1.038$ &
0.05015 & 0.0\% & 1.147\% \\
\hline
${\cal S}^3/T^\star$, $\Omega_{\hbox{\scriptsize tot}} = 1.065$ &
0.01672 & 0.0\% & 1.423\% \\
\hline
${\cal S}^3/O^\star$, $\Omega_{\hbox{\scriptsize tot}} = 1.018$ &
0.03107 & 0.0\% & 1.246\% \\
\hline
${\cal S}^3/O^\star$, $\Omega_{\hbox{\scriptsize tot}} = 1.038$ &
0.00952 & 0.0\% & 1.376\% \\
\hline
${\cal S}^3/O^\star$, $\Omega_{\hbox{\scriptsize tot}} = 1.065$ &
0.01836 & 0.0\% & 1.397\% \\
\hline
${\cal S}^3/I^\star$, $\Omega_{\hbox{\scriptsize tot}} = 1.018$ &
0.00347 & 18.0\% & 1.527\% \\
\hline
${\cal S}^3/I^\star$, $\Omega_{\hbox{\scriptsize tot}} = 1.038$ &
0.00237 & 62.8\% & 1.552\% \\
\hline
${\cal S}^3/I^\star$, $\Omega_{\hbox{\scriptsize tot}} = 1.065$ &
0.00222 & 70.7\% & 1.488\% \\
\hline
\end{tabular}
\end{minipage}
\end{table*}


\begin{figure}[htb]
\begin{center}
\vspace*{-70pt}
\hspace*{0pt}\begin{minipage}{14cm}
\includegraphics[width=9.0cm]{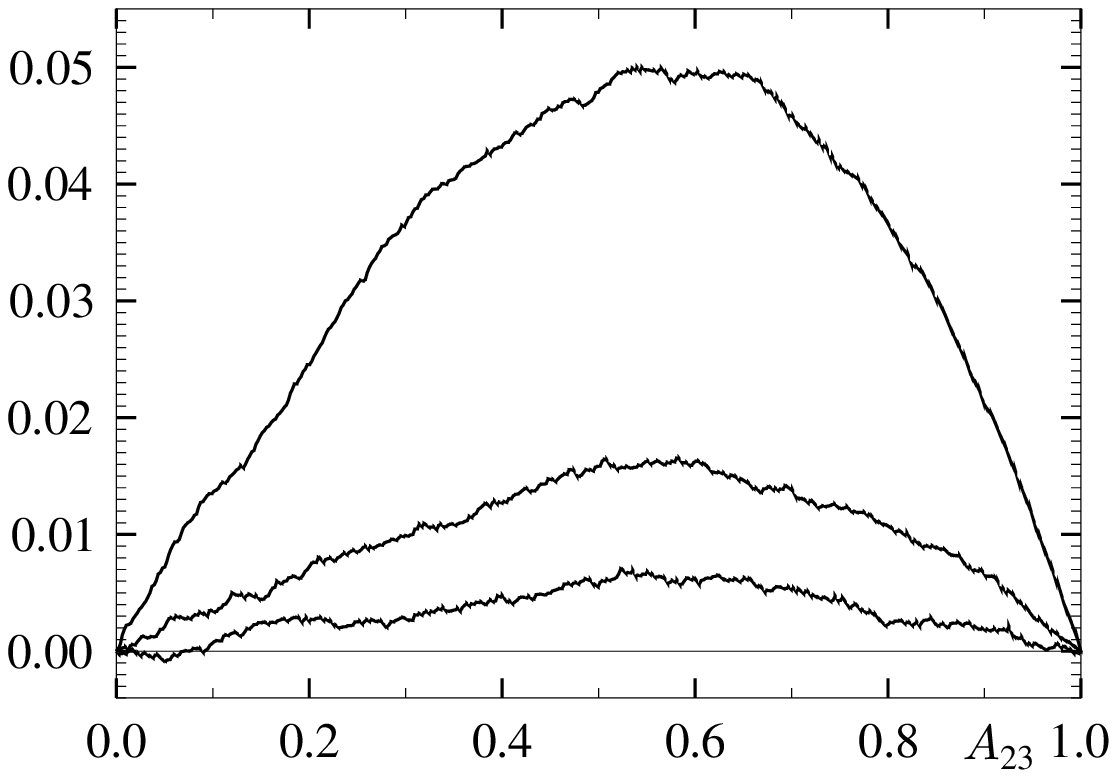}
\end{minipage}
\end{center}
\vspace*{-40pt}
\caption{\label{Fig:A23_Tetraeder}
The deviation of the cumulative distribution $P(X<A_{23})$
from the uniform distribution is shown for the 
binary tetrahedron ${\cal S}^3/{\cal T}^\star$ for
$\Omega_{\hbox{\scriptsize tot}} = 1.018$, 1.038, and 1.065.
}
\end{figure}



\begin{figure}[htb]
\begin{center}
\vspace*{-70pt}
\hspace*{0pt}\begin{minipage}{14cm}
\includegraphics[width=9.0cm]{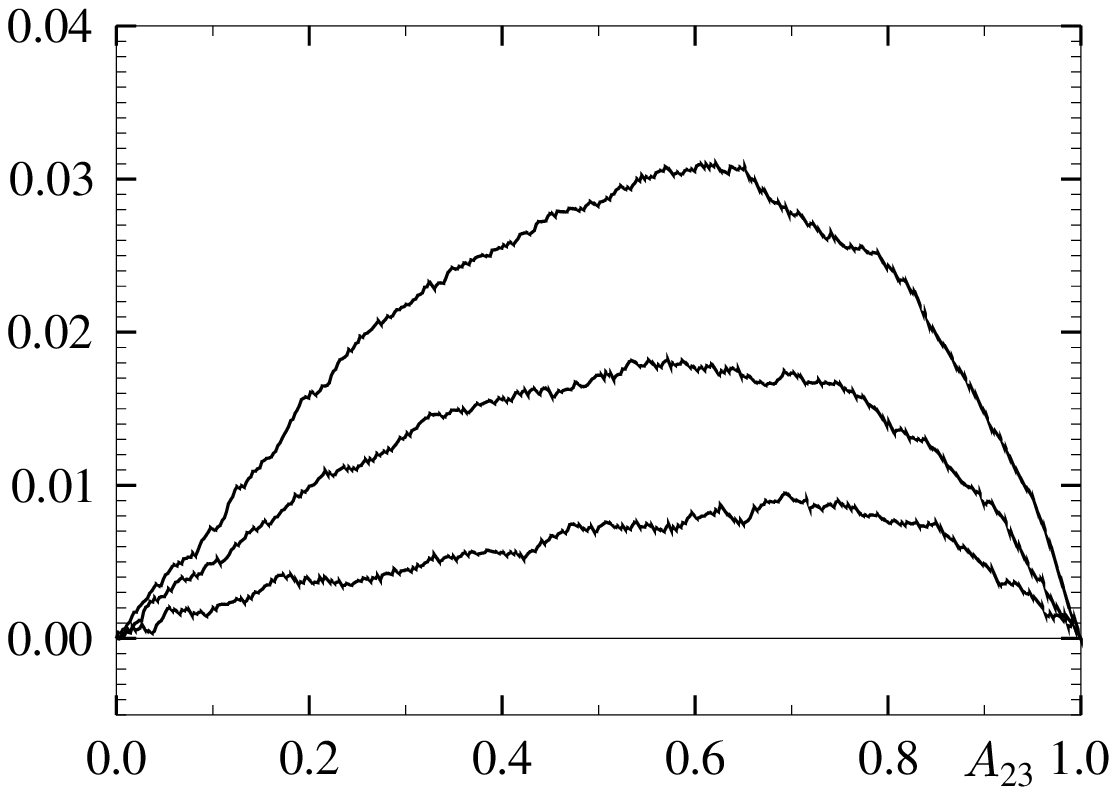}
\end{minipage}
\end{center}
\vspace*{-40pt}
\caption{\label{Fig:A23_Oktaeder}
The deviation of the cumulative distribution $P(X<A_{23})$
from the uniform distribution is shown for the 
binary octahedron ${\cal S}^3/{\cal O}^\star$ for
$\Omega_{\hbox{\scriptsize tot}} = 1.018$, 1.038, and 1.065.
}
\end{figure}



\begin{figure}[htb]
\begin{center}
\vspace*{-70pt}
\hspace*{0pt}\begin{minipage}{14cm}
\includegraphics[width=9.0cm]{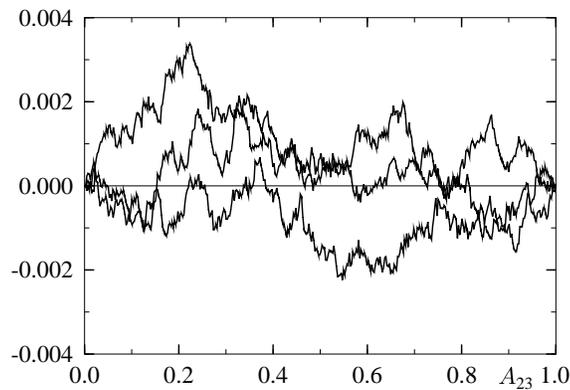}
\end{minipage}
\end{center}
\vspace*{-40pt}
\caption{\label{Fig:A23_Dodekaeder}
The deviation of the cumulative distribution $P(X<A_{23})$
from the uniform distribution is shown for the 
Poincar\'e dodecahedron ${\cal S}^3/{\cal I}^\star$ for
$\Omega_{\hbox{\scriptsize tot}} = 1.018$, 1.038, and 1.065.
}
\end{figure}


The results are shown in table \ref{Tab:A23_Spherical} and in
figures \ref{Fig:A23_Tetraeder} to \ref{Fig:A23_Dodekaeder}.
The eigenfunctions of the Poincar\'e dodecahedron ${\cal S}^3/I^\star$
do not fulfil the property i).
Although it possesses a large gap in the eigenvalue spectrum
between $n=1$ and $n=2$,
which would facilitate to satisfy property ii),
the sky maps are without property i) not aligned for this space form
as inferred from table \ref{Tab:A23_Spherical} and
figure \ref{Fig:A23_Dodekaeder}.
This is in contrast to the other two spherical space forms,
i.\,e.\ ${\cal S}^3/T^\star$ and ${\cal S}^3/O^\star$.
Here the deviations are so large such that the
Kolmogorov-Smirnov test yields almost zero probabilities $P_{KS}$.
However, these two space forms possess an anti-alignment
leading to a smaller probability for $P(A_{23} > 0.9849)$
than a uniform distribution, see table \ref{Tab:A23_Spherical}.
Thus, these spherical space forms cannot explain the quadrupole-octopole
alignment as a topological effect.

\subsection{An inhomogeneous model with negative spatial curvature}

\begin{figure}[htb]
\begin{center}
\vspace*{-70pt}
\hspace*{-90pt}\begin{minipage}{14cm}
\begin{minipage}{6cm}
\includegraphics[width=9.0cm]{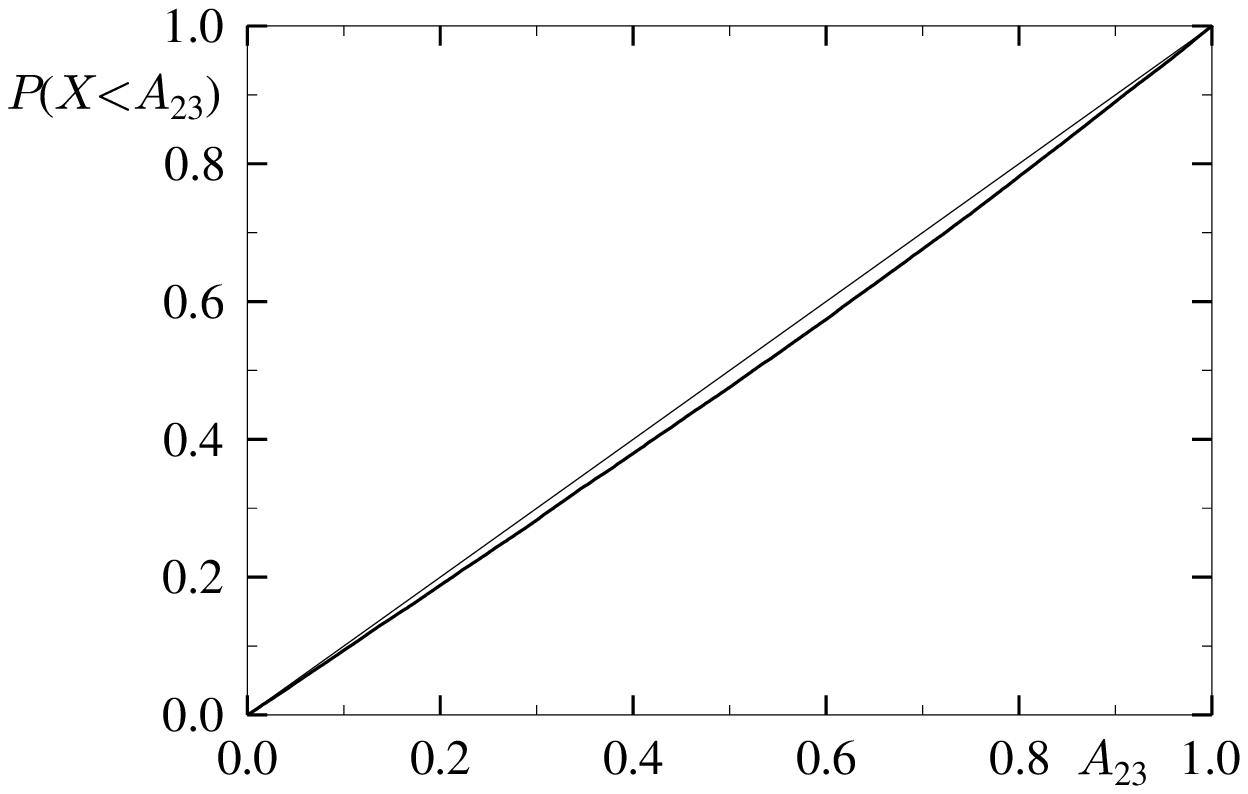}
\put(-180,155){(a)}
\put(-110,60){cusp modes}
\end{minipage}
\begin{minipage}{6cm}
\hspace*{60pt}\includegraphics[width=9.0cm]{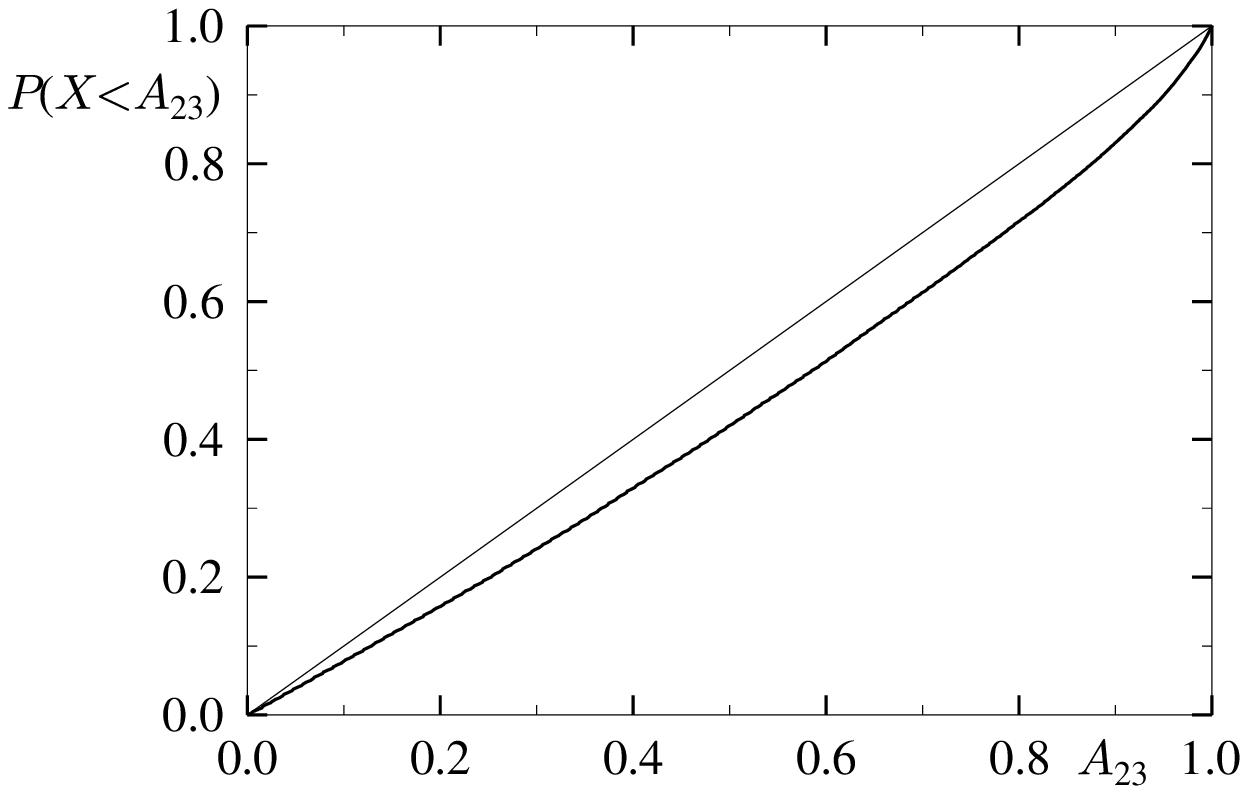}
\put(-175,155){(b)}
\put(-110,60){Eisenstein modes}
\end{minipage}
\end{minipage}
\hspace*{-90pt}\begin{minipage}{14cm}
\vspace*{-90pt}\begin{minipage}{6cm}
\includegraphics[width=9.0cm]{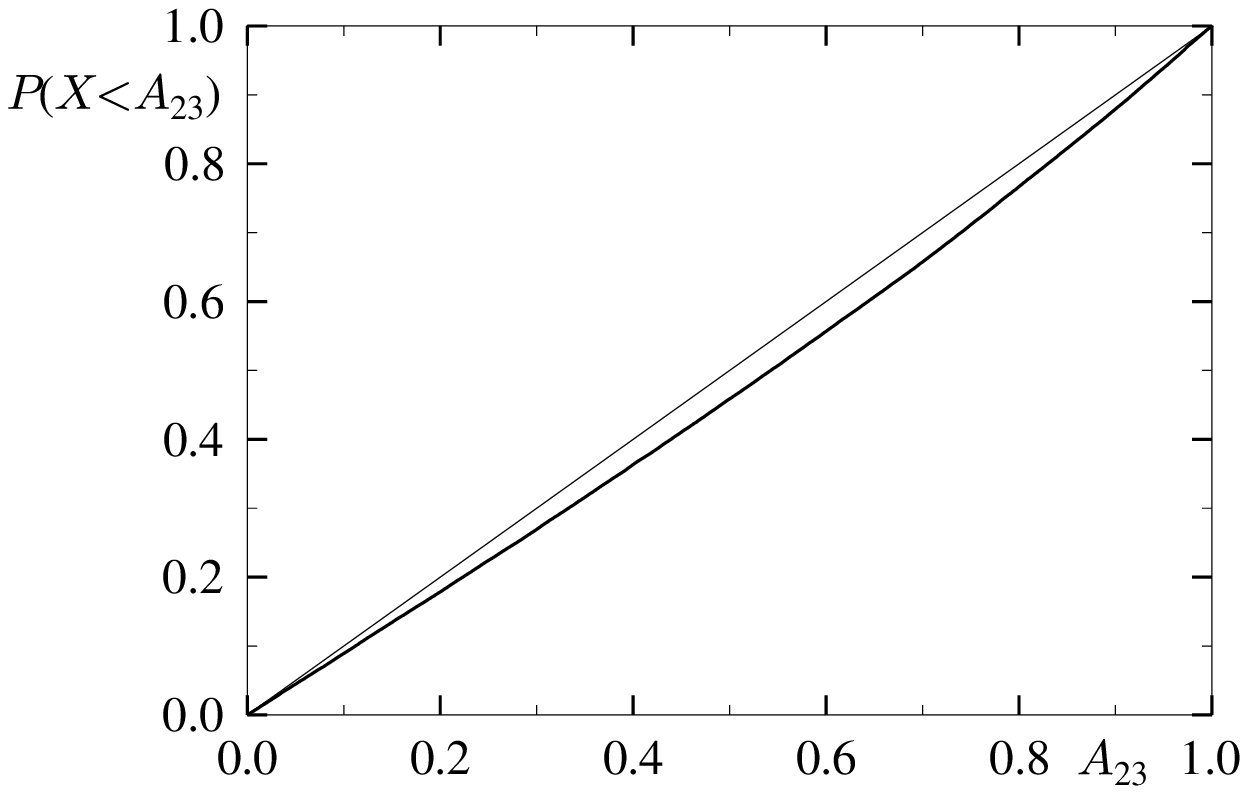}
\put(-180,155){(c)}
\put(-110,60){cusp modes}
\end{minipage}
\begin{minipage}{6cm}
\hspace*{60pt}\includegraphics[width=9.0cm]{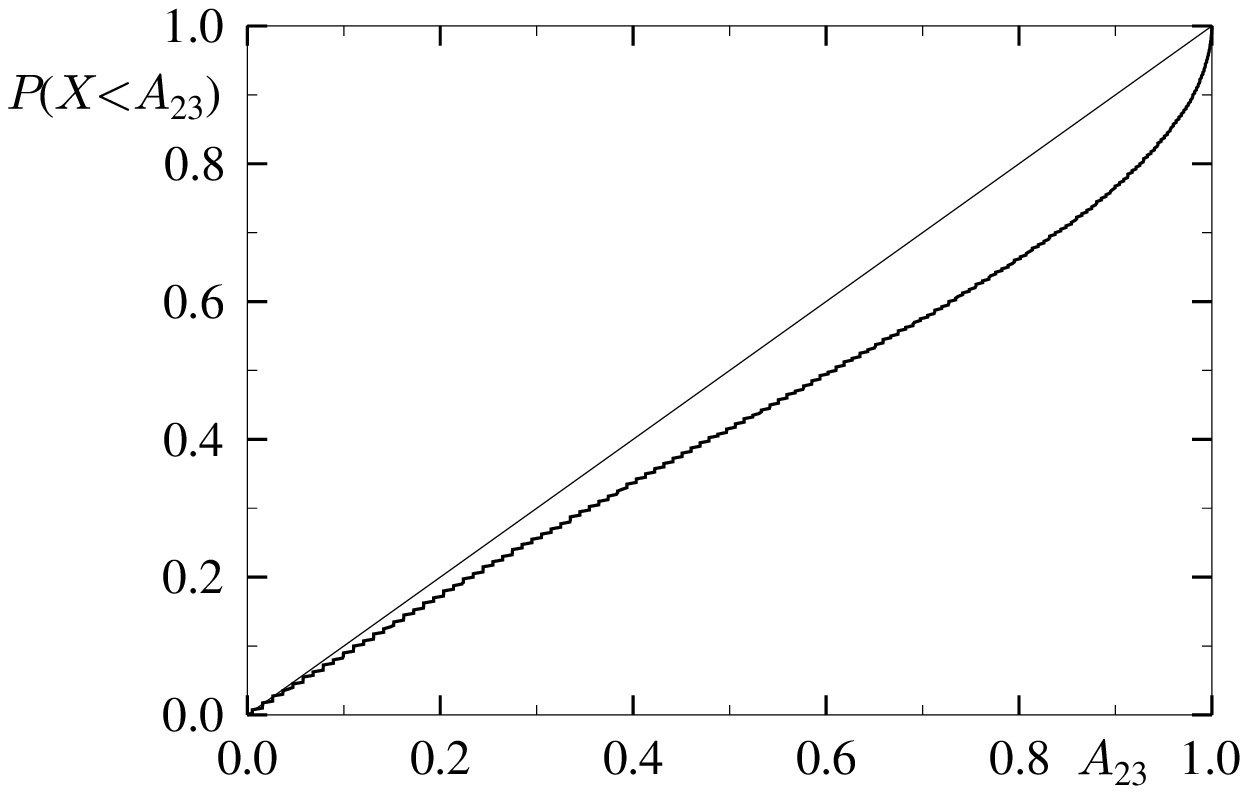}
\put(-175,155){(d)}
\put(-110,60){Eisenstein modes}
\end{minipage}
\end{minipage}
\end{center}
\vspace*{-40pt}
\caption{\label{Fig:A23_Picard}
The cumulative distribution $P(X<A_{23})$ is shown for 100\,000 realizations
of the Picard space with $\Omega_{\hbox{\scriptsize m}} =
\Omega_{\hbox{\scriptsize bar}}+\Omega_{\hbox{\scriptsize cdm}}=0.3$
and $\Omega_\Lambda = 0.65$.
Panels (a) and (b) show the results for the ``lower'' observer point,
whereas in panels (c) and (d) the observer point is high up in the horn.
}
\end{figure}

In the case of the hyperbolic Picard topology
\cite{Aurich_Lustig_Steiner_Then_2004a,Aurich_Lustig_Steiner_Then_2004b}
we show the cumulative distribution $P(X<A_{23})$
for a nearly flat model with $\Omega_{\hbox{\scriptsize tot}} = 0.95$
($\Omega_{\hbox{\scriptsize m}} =
\Omega_{\hbox{\scriptsize bar}}+\Omega_{\hbox{\scriptsize cdm}}=0.3$,
$\Omega_\Lambda = 0.65$, and $h=0.7$).
Because of the larger deviations from the uniform distribution,
$P(X<A_{23})$ is shown in figure \ref{Fig:A23_Picard}.
The Kolmogorov-Smirnov test gives almost zero probabilities $P_{KS}$
compared to the uniform distribution.
Thus, this inhomogeneous space form clearly shows a non-uniform distribution
of $A_{23}$ in contrast to the homogeneous space forms considered before.
One obtains
1.692\%, 3.818\%, 1.768\%, and 8.856\% out of 100\,000 realizations
possessing values of $A_{23}$ larger than $A_{23} = 0.9849$ for
the observer $A$ far away from the horn
using either the cusp modes or the Eisenstein modes,
and for the observer $B$ in the horn using again either cusp or
Eisenstein modes, respectively.
Thus, while the cusp modes yield a small correlation between the quadrupole
and the octopole,
the Eisenstein modes display a clear tendency for a
quadrupole-octopole alignment.
However, as discussed in \cite{Aurich_Lustig_Steiner_Then_2004a},
the primordial curvature perturbations are composed of both the
cusp and the Eisenstein modes.
Because of their different nature,
the former belonging to the discrete spectrum and being square integrable
and the latter belonging to the continuum being normalizable only to a
Dirac-$\delta$ distribution, there is no hint of their relative
contribution to the primordial perturbations.
A superposition of both types of modes would thus diminish the
alignment effect of the Eisenstein modes.
However, one has found here a topological space form
which has naturally a higher quadrupole-octopole alignment.

%
%

\section{Multipole vectors in multi-connected spaces}

The large-scale CMB anomalies have also been studied
with respect to the Maxwell multipole vectors for the low multipoles.
Let us introduce the corresponding quantities
for which we will present the statistical properties
for multi-connected space forms.
From the area vectors (\ref{Eq:Area_vectors})
the following scalar products can be formed
\cite{Schwarz_Starkman_Huterer_Copi_2004}
for $l=2$ and $l=3$
\begin{eqnarray}
\label{Eq:Area_vector_products}
\tilde A_1 & = & \nonumber
|\, \vec w^{(2,1,2)} \,\cdot\, \vec w^{(3,1,2)}\, |
\\
\tilde A_2 & = &
|\, \vec w^{(2,1,2)} \,\cdot\, \vec w^{(3,1,3)}\, |
\\
\tilde A_3 & = & \nonumber
|\, \vec w^{(2,1,2)} \,\cdot\, \vec w^{(3,2,3)}\, |
\hspace{10pt}.
\end{eqnarray}
Ordered with respect to their magnitudes,
the scalar products are denoted by $A_1$, $A_2$ and $A_3$,
i.\,e.\  $A_1>A_2>A_3$.
The analogous scalar products $D_1$, $D_2$ and $D_3$ are obtained by using
normalised vectors $\vec w^{(l,i,j)}/|\vec w^{(l,i,j)}|$ instead
of the area vectors $\vec w^{(l,i,j)}$
in (\ref{Eq:Area_vector_products}).
From these dot products the following statistical measures can be constructed
\begin{equation}
\label{Eq:S_m_statistic}
S_m \; := \; \frac 1m \sum_{i=1}^m A_i
\hspace{10pt} , \hspace{10pt}
m = 1,2,3
\end{equation}
and
\begin{equation}
\label{Eq:T_m_statistic}
T_m \; := \; 1 - \frac 1m \sum_{i=1}^m (1-A_i)^2
\hspace{10pt} , \hspace{10pt}
m = 1,2,3
\hspace{10pt} .
\end{equation}

%
%
\begin{figure}[htb]
\begin{center}
\vspace*{-70pt}
\hspace*{-90pt}\begin{minipage}{14cm}
\begin{minipage}{6cm}
\includegraphics[width=9.0cm]{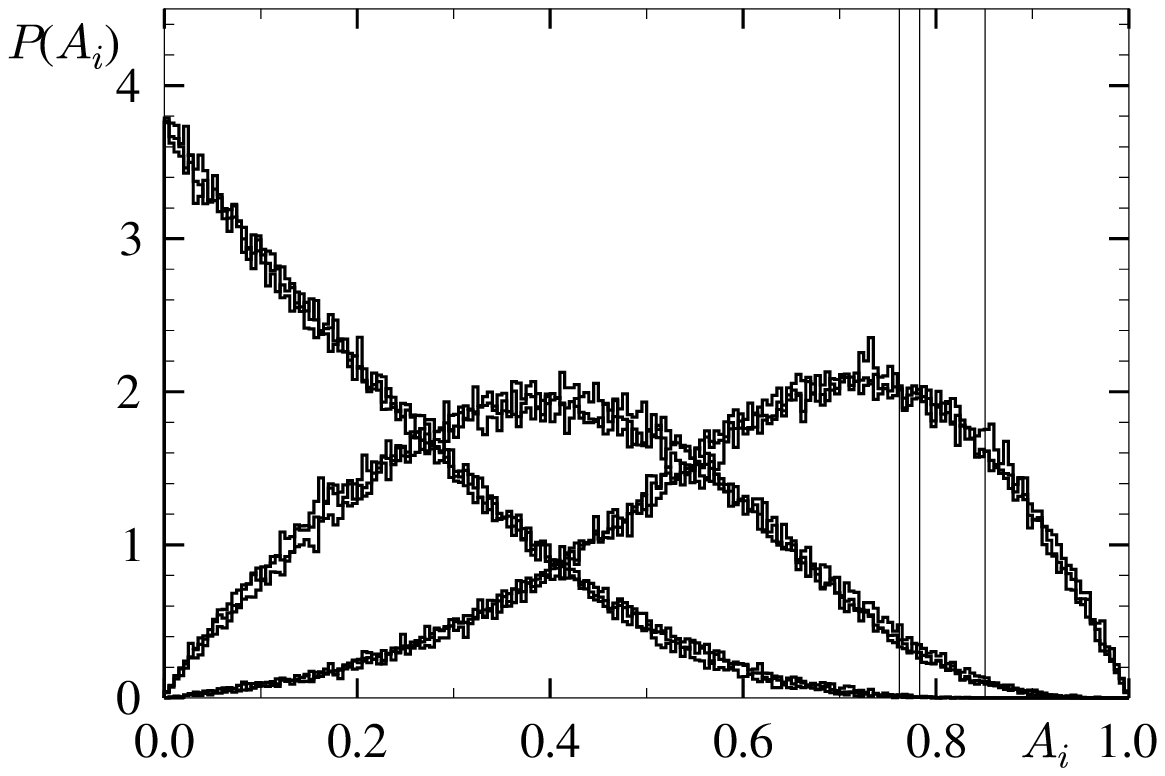}
\put(-170,155){(a)}
\end{minipage}
\begin{minipage}{6cm}
\hspace*{60pt}\includegraphics[width=9.0cm]{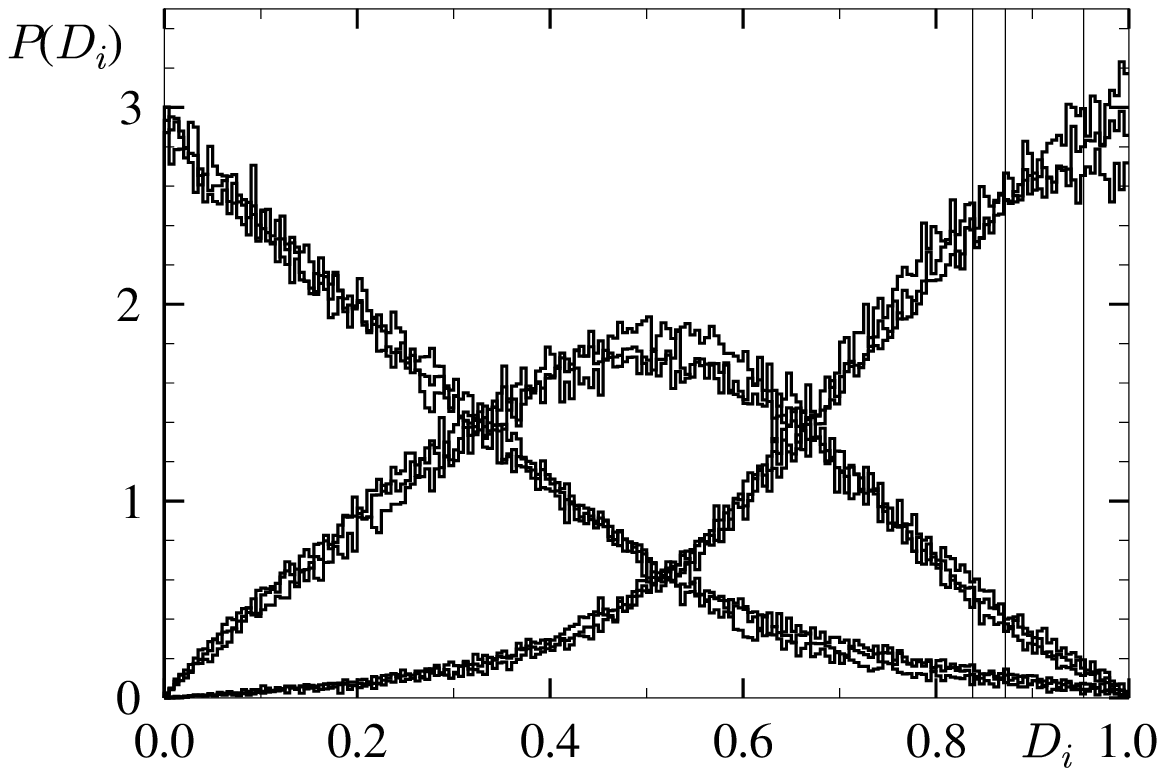}
\put(-175,155){(b)}
\end{minipage}
\end{minipage}
\hspace*{-90pt}\begin{minipage}{14cm}
\vspace*{-90pt}\begin{minipage}{6cm}
\includegraphics[width=9.0cm]{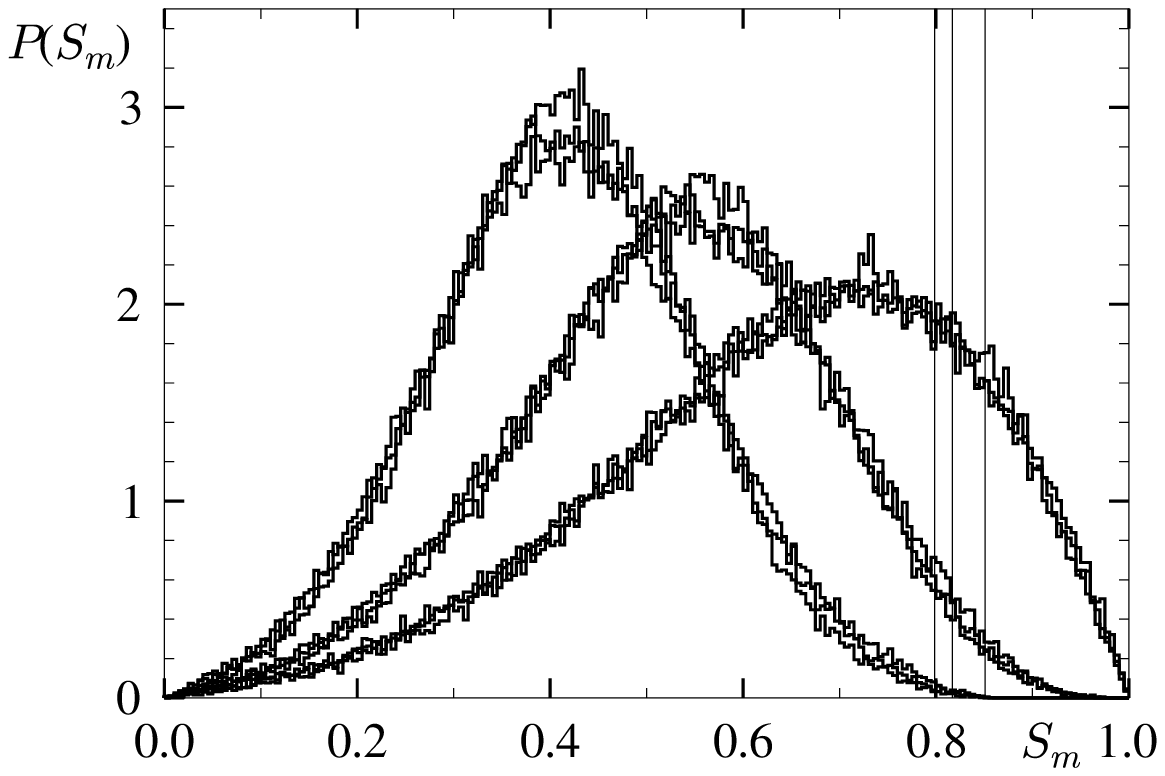}
\put(-180,155){(c)}
\end{minipage}
\begin{minipage}{6cm}
\hspace*{60pt}\includegraphics[width=9.0cm]{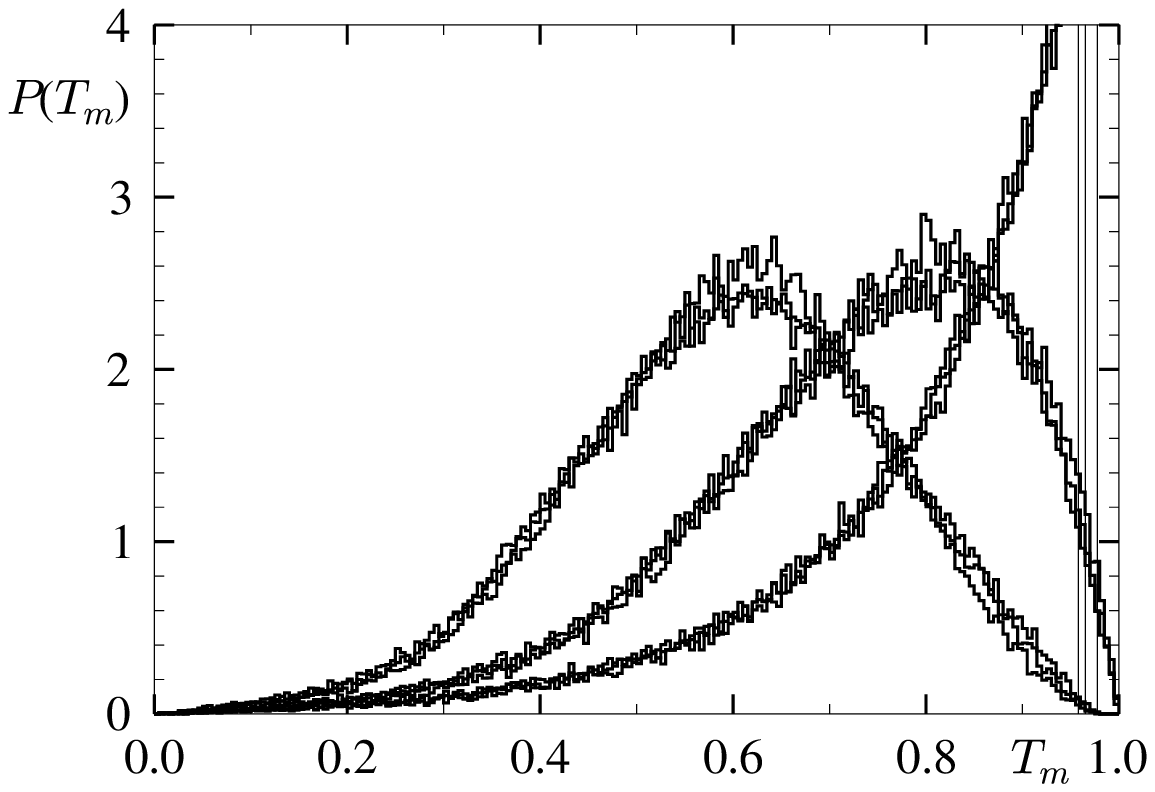}
\put(-175,155){(d)}
\end{minipage}
\end{minipage}
\end{center}
\vspace*{-40pt}
\caption{\label{Fig:MP_Multi_conntected}
The distributions of $A_i$ (panel a), $D_i$ (panel b), $S_m$ (panel c),
and $T_m$ (panel d) with $i,m=1,2,3$
are presented for the torus with side length $L=1.0$,
the dodecahedron ${\cal S}^3/I^\star$ for
$\Omega_{\hbox{\scriptsize tot}} = 1.018$,
and the Picard topology using the cusp modes
for the observer position $A$.
The histograms are based on $10^5$ sky realizations.
The vertical lines indicate the corresponding WMAP values obtained
from the TdOH sky map.
}
\end{figure}

There are several estimates for $A_i$ and $D_i$ obtained from
the WMAP observations based on the different maps derived from the data
\cite{Copi_Huterer_Schwarz_Starkman_2005,Copi_Huterer_Schwarz_Starkman_2006}.
The change of the quadrupole in the ILC 3yr map
(see e.\,g. fig.\ 7 in \cite{deOliveira-Costa_Tegmark_2006})
has caused slightly different values in the 3 year data
\cite{Copi_Huterer_Schwarz_Starkman_2006}
but all give surprisingly high values.
In figure \ref{Fig:MP_Multi_conntected} we show the corresponding values
as vertical lines obtained from the TdOH map
\cite{Tegmark_deOliveira_Costa_Hamilton_2003}
\begin{equation}
\label{Eq:A_i}
A_1 \; = \; 0.851
\hspace{10pt} , \hspace{10pt}
A_2 \; = \; 0.783
\hspace{10pt} , \hspace{10pt}
A_3 \; = \; 0.762
\hspace{10pt} ,
\end{equation}
and
\begin{equation}
\label{Eq:D_i}
D_1 \; = \; 0.953
\hspace{10pt} , \hspace{10pt}
D_2 \; = \; 0.872
\hspace{10pt} , \hspace{10pt}
D_3 \; = \; 0.838
\hspace{10pt} .
\end{equation}
From these values the values for $S_m$ and $T_m$ follow by
(\ref{Eq:S_m_statistic}) and (\ref{Eq:T_m_statistic}).
The alignment causes large values for $A_i$, $D_i$, $S_m$, and $T_m$
for $i,m=3$, which are untypical for a ``generic'' sky map.
In figure \ref{Fig:MP_Multi_conntected} the distributions for
these quantities are shown obtained from $10^5$ sky realizations
for the following three topological spaces:
the torus with side length $L=1.0$,
the dodecahedral space ${\cal S}^3/I^\star$ with
$\Omega_{\hbox{\scriptsize tot}} = 1.018$,
and the Picard topology using the cusp modes
for the observer position $A$ for $\Omega_{\hbox{\scriptsize tot}} = 0.95$.
All models are plotted onto the same corresponding graph
since they give very similar distributions
which are in turn similar to simply-connected space forms.

In panel (a) the distributions for $A_i$, $i=1,2,3$ are shown.
With increasing value of $i$, the maxima of the histograms shift to
smaller values, of course.
The observed WMAP value for $A_1$ is not unusual as a comparison with
the corresponding histograms reveals.
The value of $A_3$, however, has only a marginal overlap with the
tail of the corresponding histograms for all three considered topologies.
The same behaviour is observed for the normalized scalar products $D_i$
shown in panel (b).
In \cite{Weeks_Gundermann_2006} the distribution of $D_i$ is studied
simultaneously for $i=1,2,3$ for the dodecahedral space ${\cal S}^3/I^\star$
and found to be equidistributed,
i.\,e.\ if all three histograms in our panel (b) would be added.
This is in agreement with our results.
We applied the Kolmogorov-Smirnov test which shows for the
dodecahedral space ${\cal S}^3/I^\star$ an equidistribution.
The other two spherical space forms yield zero probabilities for that
in agreement with the above results.

In panels (c) and (d) the distributions for $S_m$ and $T_m$,
eqs.\,(\ref{Eq:S_m_statistic}) and (\ref{Eq:T_m_statistic}), are shown.
Again the distributions with smaller values at their maxima belong to
higher values of $m$.
As it is the case for $A_3$ and $D_3$, there is only a very small overlap
with the tail of the distributions for $S_3$ and $T_3$
with the corresponding observed values.

Since the four distributions are similar for the simply-connected
universes and the considered multi-connected space forms,
the non-trivial topology does not help to explain the anomalous
quadrupole and octopole properties.
But that does not disfavour multi-connected space forms,
since the simply-connected alternatives do not yield higher
probabilities to remedy the anomalous behaviour.
As discussed in the introduction there are several suggestions
that the anomalous quadrupole and octopole properties are not of
cosmological origin.
In that case no explanation would be required from a satisfactory
cosmological concordance model.

\section{Conclusion}

We have investigated the question whether the strange alignment
observed in low CMB multipoles can be explained by multi-connected space forms.
There are several examples of such spaces
which can explain the missing anisotropy power at large angular scales
as measured by the temperature correlation function $C(\vartheta)$,
$\vartheta > 60^\circ$, or the angular power spectrum $C_l$ for $l=2,3$.
It is thus natural to ask for the alignment properties in such spaces.

It is shown that two requirements are necessary for a non-trivial topology
to be able to generate the observed alignment.
The eigenfunctions belonging to the smallest eigenvalue have to obey
a strong topological alignment, i.\,e.\ they should already have an aligned
quadrupole and octopole.
In order to make the survival of this property possible in CMB sky maps,
it is necessary that the transfer function $T_l(k)$ gives those eigenfunctions
with a topological alignment a sufficiently strong weight.
In the case of the hypertorus we demonstrate this connection explicitly.
It is found that depending on the chosen side length of the hypertorus
the CMB sky maps are slightly aligned in some cases and in others not. 
The alignment is, however, not strong enough in order to state
that the hypertorus model could solve the alignment riddle.
In addition, three spherical models are considered of which one,
the dodecahedron, shows no alignment whereas the other two,
the binary tetrahedron and the binary octahedron, display anti-alignment
which reduces the probability in comparison to isotropic models.
Furthermore, as an example of an inhomogeneous space form,
the Picard model of hyperbolic space is studied.
This model provides an example of a space form possessing an intrinsic
alignment which increases the probability compared to isotropic models.

But in no case we have found a model,
where a significant fraction of the simulations possesses the alignment
observed in the CMB sky.
It remains to be seen whether there are other space forms
which can more easily explain the alignment than the space forms considered
in this paper.


\section*{Acknowledgment}

\vspace*{-3pt}
The computation of the multipole vectors $\hat v^{(l,j)}$ has been
carried out with the algorithm kindly provided by
\cite{Copi_Huterer_Starkman_2004}.
\vspace*{-3pt}


\section*{References}

\bibliography{../bib_chaos,../bib_astro}
\bibliographystyle{h-physrev3}

\end{document}